\pdfoutput=1
\newcommand*{\ATLASLATEXPATH}{latex/}
\documentclass[UKenglish,texlive=2016,cernpreprint]{\ATLASLATEXPATH atlasdoc}

\usepackage[centering,scale=0.75]{geometry}

\usepackage[subfigure,backend=bibtex]{\ATLASLATEXPATH atlaspackage}
\usepackage{longtable}
\usepackage{verbatim}
\usepackage[maxbibnames=4]{\ATLASLATEXPATH atlasbiblatex}

\usepackage{\ATLASLATEXPATH atlascontribute}

\usepackage{\ATLASLATEXPATH atlasphysics}

\graphicspath{{logos/}{figures/}{../INT_XS_ljets_8TeVWup/}}

\usepackage{TOPQ-2016-08-defs}

\AtlasTitle{Measurement of the inclusive and fiducial \ttbar
production cross-sections in the lepton+jets channel in $pp$ collisions at
$\sqrt{s} = \SI{8}{\TeV}$ with the ATLAS detector}

\author{The ATLAS Collaboration}

\AtlasRefCode{TOPQ-2016-08}

\PreprintIdNumber{CERN-EP-2017-276}

\AtlasJournal{\EPJC}
\AtlasJournalRef{Eur.\ Phys.\ J.\ C 78 (2018) 487}
\AtlasDOI{10.1140/epjc/s10052-018-5904-z}

\AtlasAbstract{%

The inclusive and fiducial \ttbar production cross-sections are measured in the
lepton+jets channel using \SI{20.2}{\per\fb} of proton--proton collision data at
a centre-of-mass energy of \SI{8}{\TeV} recorded with the ATLAS detector at the
LHC.
Major systematic uncertainties due to the modelling of the jet energy scale and
$b$-tagging efficiency are constrained by separating selected events into three
disjoint regions. In order to reduce systematic uncertainties in the most
important background, the \Wjets process is modelled using \Zjets events in a
data-driven approach.
The inclusive \ttbar cross-section is measured with a precision of
\SI{5.7}{\percent} to be $\sigma_{\text{inc}}(\ttbar) = 248.3 \pm 0.7 \,
(\mathrm{stat.})  \pm 13.4 \, (\mathrm{syst.}) \pm 4.7 \,
(\mathrm{lumi.})~\text{pb}$, assuming a top-quark mass of \SI{172.5}{\gev}. The
result is in agreement with the Standard Model prediction. The cross-section is
also measured in a phase space close to that of the selected data. The fiducial
cross-section is $\sigma_{\text{fid}}(\ttbar) = 48.8 \pm 0.1 \, (\mathrm{stat.})  \pm 2.0 \,
(\mathrm{syst.}) \pm 0.9 \, (\mathrm{lumi.})~\text{pb}$ with a
precision of \SI{4.5}{\percent}.
}

\AtlasCoverSupportingNote{INT note}{https://cds.cern.ch/record/2126101}
\AtlasCoverSupportingNote{Changes to draft1}{https://cds.cern.ch/record/2289060}
\AtlasCoverEgroupEditors{atlas-TOPQ-2016-08-editors@cern.ch}

\AtlasCoverEgroupEdBoard{atlas-TOPQ-2016-08-editorial-board@cern.ch}

\hypersetup{pdftitle={ATLAS document},pdfauthor={The ATLAS Collaboration}}

\newif\ifPrintNotes
\PrintNotesfalse
\ifPrintNotes
  \newcommand{\PMnote}[1]{\textcolor{magenta}{[PM: #1]}}
  \newcommand{\ABnote}[1]{\textcolor{cyan}{[AB: #1]}}
  \newcommand{\DHnote}[1]{\textcolor{green}{[DH: #1]}}
  \newcommand{\EBnote}[1]{\textcolor{red}{[Note for reviewer: #1]}}
\else
  \newcommand{\PMnote}[1]{\mbox{}}
  \newcommand{\ABnote}[1]{\mbox{}}
  \newcommand{\DHnote}[1]{\mbox{}}
  \newcommand{\EBnote}[1]{\mbox{}}
\fi

\begin{document}

\maketitle


\section{Introduction}
\label{sec:intro}

The top quark is the most massive known elementary particle.
Given that its Yukawa coupling to the Higgs boson is close to unity,
it may play a special role in electroweak symmetry breaking
\cite{Higgs:1964pj,Englert:1964et}.
Studies of top-quark production and decay are major research goals at the
LHC, providing both a precise probe of the Standard Model
(SM)~\cite{Baak:2014ora} and a window on physics beyond the Standard Model
(BSM)~\cite{Buckley:2015lku}.
The LHC supplies a large number of top-quark events to its major experiments,
offering an excellent environment for such studies.

In proton--proton collisions, the dominant top-quark production process is pair
production via the strong interaction.
The measurement of the production cross-section  provides a stringent test of QCD calculations with heavy
quarks~\cite{Czakon:2013goa}, allows a determination of the top-quark mass in a well-defined
renormalisation scheme~\cite{Langenfeld:2009wd,Wang:2017kyd}, and can be
sensitive to potential new physics such as top-quark partners degenerate in
mass with the SM top quark~\cite{Franceschini:2016tbl}.
 
The predicted inclusive \ttbar cross-section at a centre-of-mass energy of
$\sqrt{s} = \SI{8}{\TeV}$, assuming a top-quark mass $\mtop = \SI{172.5}{\gev}$, is
\begin{equation}
\sigma(pp \rightarrow \ttbar) \ = \ 253^{+13}_{-15}\ \
\ \mathrm{pb}.
\label{eq:theo}
\end{equation}
It is calculated at next-to-next-to-leading order (NNLO) in QCD including
resummation of next-to-next-to-leading logarithmic (NNLL) soft-gluon terms with \texttt{Top++} (v2.0)~\cite{Cacciari:2011hy,Baernreuther:2012ws,Czakon:2012zr,Czakon:2012pz,Czakon:2013goa,Czakon:2011xx}.
The QCD scale uncertainties are determined as the maximum deviation in the
predicted cross-section for the six probed variations 
following a prescription referred to as independent restricted scale
variations.
Here the renormalisation scale (\mur) and the factorisation scale
(\muf) are varied independently to half the default scale and 
twice the default scale omitting the combinations
($0.5 \murdef$, $2.0 \mufdef$) and ($2.0 \murdef$, $0.5 \mufdef$).
The uncertainties due to the parton distribution functions (PDFs) and
$\alphas$ are calculated using the PDF4LHC prescription~\cite{Botje:2011sn}
where the uncertainties of the \mstw \SI{68}{\percent} CL
NNLO~\cite{Martin:2009iq, Martin:2009bu}, \ct NNLO~\cite{Lai:2010vv, Gao:2013xoa} and
\nnpdf~\cite{Ball:2012cx} PDF sets are added in quadrature to the $\alphas$
uncertainty. Comparable results are obtained using a different resummation
technique as reported in Refs.~\cite{Muselli:2015kba,Kidonakis:2014isa}.
The predicted cross-section's total scale, PDF and $\alphas$ uncertainty of
about \SI{6}{\percent} sets the current goal for the experimental precision.

Measurements of the \ttbar cross-section have been published for several
centre-of-mass energies between \num{1.96} and \SI{13}{\TeV} in $p\bar{p}$ and
$pp$ collisions. At the Tevatron, the uncertainty in the \ttbar cross-section
measured by the D0 and CDF collaborations at a centre-of-mass
energy of \SI{1.96}{\TeV} is \SI{5.4}{\percent}~\cite{Aaltonen:2013wca}.
The most precise measurement for a centre-of-mass energy of \SI{8}{\TeV}, with a
total uncertainty of \SI{3.2}{\percent}, was performed by the ATLAS
Collaboration in the dilepton channel, where both top quarks decay via $t \to
\ell \nu b$~\cite{TOPQ-2013-04-witherratum}. The final-state charged lepton
$\ell$ is either an electron or a muon.\footnote{%
  Events involving $W\rightarrow \tau\nu$ decays with a subsequent decay of the
  $\tau$ lepton to either $e\nu_e\nu_\tau$ or $\mu\nu_\mu\nu_\tau$ are included
  in the signal.} Further measurements at \num{7}, \num{8} and
\SI{13}{\TeV} in the same final state were published by the ATLAS and CMS
collaborations~\cite{TOPQ-2015-09, CMS-TOP-13-004, CMS-TOP-16-005}.
Additionally, a measurement of the \ttbar production cross-section in the
forward region at \SI{8}{\TeV} was published by the LHCb
collaboration~\cite{Aaij:2016vsy}.

The measurement reported in this paper is performed in the
lepton+jets final state, where one $W$
boson decays leptonically and the other $W$ boson decays hadronically, i.e.
\begin{equation*}
\ttbar \ \rightarrow W^+ W^- b\bar{b} \rightarrow \ell \nu q\bar{q}' b\bar{b}.
\end{equation*}
Results are reported for both the full phase space and for a fiducial
phase space close to the selected dataset.

Since experimental uncertainties may affect each decay mode differently, it is
important to determine whether \ttbar cross-sections measured in different decay
modes are consistent with each other. Furthermore, new physics processes can
contribute in different ways to the different decay modes.

The analysis is based on data collected at a $pp$
centre-of-mass energy of $\sqrt{s} = \SI{8}{\TeV}$.
The most precise cross-section previously measured in this channel at
$\sqrt{s} = \SI{8}{\TeV}$ was published by the CMS Collaboration and reached an
uncertainty of \SI{6.8}{\percent}~\cite{CMS-TOP-12-006}.
This analysis supersedes the previous measurement from the ATLAS
Collaboration, which achieved a total uncertainty of
\SI{9.4}{\percent} using the same dataset~\cite{TOPQ-2013-06}.
This analysis improves on the previous result by splitting the overall sample of \ttbar
candidates into three signal regions and by constraining important sources
of systematic uncertainty.

\newcommand{\AtlasCoordFootnote}{%
ATLAS uses a right-handed coordinate system with its origin at the nominal
interaction point (IP) in the centre of the detector and the $z$-axis along the
beam pipe.
The $x$-axis points from the IP to the centre of the LHC ring, and the $y$-axis
points upwards.
Cylindrical coordinates $(r,\phi)$ are used in the transverse plane, $\phi$
being the azimuthal angle around the $z$-axis.
The pseudorapidity is defined in terms of the polar angle $\theta$ as $\eta =
-\ln \tan(\theta/2)$, and the distance $\Delta R$ in the $\eta$--$\phi$ space
is defined as $\Delta R \equiv \sqrt{(\Delta\eta)^{2} + (\Delta\phi)^{2}}$.}

\section{ATLAS detector}
\label{sec:atlas}

The ATLAS detector~\cite{PERF-2007-01} is a multi-purpose particle detector with
forward-backward symmetry and a cylindrical
geometry.\footnote{\AtlasCoordFootnote}
The inner detector (ID) tracking system is surrounded by a thin
superconducting solenoid magnet, electromagnetic and hadronic calorimeters, and a muon
spectrometer (MS) in a magnetic field generated by three superconducting
toroidal magnets of eight coils each.
The inner detector, in combination with the \SI{2}{\tesla} magnetic field from
the solenoid, provides precision momentum measurements for charged particles within the
pseudorapidity range $\vert \eta \vert <$ 2.5. It consists of, from the
interaction point to the outside,  a silicon pixel detector and a silicon
microstrip detector (together allowing a precise and efficient identification of
secondary vertices), complemented with a straw-tube tracker contributing
transition radiation measurements to electron identification. The calorimeter
system covers the pseudorapidity range $\vert \eta \vert <$ 4.9. A
high-granularity liquid-argon (LAr) sampling calorimeter with lead absorbers
provides the measurement of electromagnetic showers within $\vert \eta \vert <$
3.2.
In the ID acceptance region, $\vert \eta \vert <$ 2.5, the innermost layer has a
fine segmentation in $\eta$ to allow separation of electrons and photons from
$\pi^0$ decays and to improve the resolution of the shower position and
direction measurements.
Hadronic showers are measured by a steel/plastic-scintillator tile calorimeter
in the central region,  $\vert \eta \vert <$ 1.7, and by a LAr calorimeter in
the endcap region, 1.5 $<\vert \eta \vert <$ 3.2. In the forward region,
measurements of electromagnetic and hadronic showers are provided by a LAr
calorimeter covering the pseudorapidity range 3.1 $<\vert \eta \vert <$ 4.9. The
MS combines trigger and high-precision tracking detectors, and allows
measurements of charged-particle trajectories within $\vert \eta \vert <$ 2.7.
The combination of all ATLAS detector subsystems provides charged-particle
tracking, along with identification for charged leptons and photons, in the
pseudorapidity range $\vert \eta \vert <$ 2.5.

A three-level trigger system is used to select interesting
events~\cite{PERF-2011-02}.
A hardware-based first-level trigger uses a subset of detector
information to bring the event rate below \SI{75}{\kHz}.
Two additional software-based trigger levels together reduce 
the event rate to about \SI{400}{\Hz} on average, depending on the data-taking
conditions.

\section{Data and simulated events}
\label{sec:samples}

This analysis is performed using $pp$ collision data 
recorded at a centre-of-mass energy of $\sqrt{s}=\SI{8}{\TeV}$, corresponding
to the full 2012 dataset.
The data-taking periods in which all the subdetectors were operational are
considered, resulting in a data sample with an integrated luminosity of $\cal{L}_\mathrm{int} =
\SI{20.2}{\per\fb}$.

Detector and trigger simulations were performed 
within the {\textsc GEANT4} framework~\cite{SOFT-2010-01,Agostinelli:2002hh}.
The same offline reconstruction methods 
used on data are applied to the simulated events.
Minimum-bias events generated with \PYTHIAV{8}\cite{Sjostrand:2014zea} were
used to simulate multiple $pp$ interactions (pile-up).
The distribution of the number of pile-up interactions in the simulation 
is reweighted according to the instantaneous luminosity spectrum in the data.

Signal \ttbar events were simulated using the \POWHEGBOX event generator
(r3026)~\cite{Alioli:2010xd,Frixione:2007nw} with the \ct PDF
set~\cite{Lai:2010vv}.
The renormalisation and factorisation scales in the matrix element calculation
were set to the value $\mu=\sqrt{\mtop^2 + \pT^2(t)}$ where $\pT(t)$ is the
top-quark transverse momentum, evaluated for the underlying Born configuration,
i.e. before radiation.
The \hdamp parameter, which controls the transverse momentum, \pT, of the first
emission beyond the Born configuration, was set to \mtop. The main effect
of this is to regulate the high-\pT gluon emission against which the \ttbar
system recoils.
Parton shower (PS), hadronisation and the underlying event were simulated with
\PYTHIA~(v6.428)~\cite{Sjostrand:2006za} and the Perugia2011C set of tuned
parameters~\cite{Skands:2010ak}.

For systematic studies of the \ttbar process, alternative event generators and
variations of the tuned parameter values in \PYTHIA are used.
The \POWHEGBOX event generator, using the same configuration as the nominal
sample, interfaced to \HERWIG (v6.5.20)~\cite{Corcella:2000bw} is used for hadronisation-modelling studies, while \MCatNLO
(v4.09)~\cite{Frixione:2002ik,Frixione:2003ei} interfaced to \HERWIG is used to
study the dependence on the matching method between the next-to-leading-order
(NLO) matrix element (ME) generation and the PS evolution.
In the case of events showered by \HERWIG, the
\JIMMY~(v4.31)~\cite{Butterworth:1996zw} model with the ATLAS
AUET2~\cite{ATL-PHYS-PUB-2011-008} set of tuned parameters
were used to simulate the underlying event.
Variations of the amount of additional radiation are studied using events
generated with the \POWHEGBOX + \PYTHIA event generators after changing the
scales in the ME and the scales in the parton shower simultaneously.
In these samples, a variation of the factorisation and renormalisation 
scales by a factor of 2 was combined with the Perugia2012radLo parameters
and a variation of both scale parameters by a factor of 0.5 was combined with
the Perugia2012radHi parameters~\cite{Skands:2010ak}.
In the second case, the $\hdamp$ parameter was also changed
and set to twice the top-quark mass~\cite{ATL-PHYS-PUB-2015-002}.

The associated production of an on-shell $W$ boson and a top quark ($Wt$), and
single top-quark production in the $s$- and $t$-channel, were simulated by the
\POWHEGBOX(r2819, r2556) event
generator~\cite{Alioli:2009je,Re:2010bp} with the \ct PDF set
interfaced to \PYTHIA using the Perugia2011C set of tuned parameters. The $Wt$
process has a predicted production cross-section of
\SI{22.3}{pb}~\cite{Kidonakis:2010ux}, calculated to approximate NNLO accuracy
with an uncertainty of \SI{7.6}{\percent} including scale and PDF uncertainties.
The cross-sections for single top-quark production in the $s$- and $t$-channel
are calculated with the Hathor~v2.1 tool~\cite{Kant:2014oha} to NLO precision,
based on work documented in Ref.~\cite{Campbell:2009ss}.
Uncertainties from variations of scales used in the ME and PDFs are estimated
using the same methodology as for \ttbar production. For $t$-channel production, this
leads to a cross-section of \SI{84.6}{\pb} with a total uncertainty of
\SI{4.6}{\percent}, while for $s$-channel production a cross-section of
\SI{5.2}{pb} with a total uncertainty of \SI{4.2}{\percent} is predicted.

All top-quark processes were simulated with a top-quark mass of \SI{172.5}{\GeV}
and a width of \SI{1.32}{\GeV} modelled using a Breit--Wigner distribution.
The top quark is assumed to decay via $t \to Wb$ \SI{100}{\percent} of the
time.

Vector-boson production in association with jets ($W$/$Z$+jets) was simulated
with \ALPGEN~(v2.14)~\cite{Mangano:2002ea}, using the CTEQ6L1 set of PDFs~\cite{Pumplin:2002vw}.
The partonic events were showered with \PYTHIA using the Perugia2011C set of
tuned parameters. Simulated \Wjets, $W+b\bar{b}, W+c\bar{c}, W+c$ and
$Z$+jets, $Z+b\bar{b}, Z+c\bar{c}$ events with up to five additional partons
were produced, and the overlap between the ME and the PS was removed with the ``MLM'' matching scheme \cite{Alwall:2007fs}.
The double-counting between the inclusive $W+n$ parton samples and
dedicated samples with at least one heavy quark ($c$- or
$b$-quark) in the ME was removed by vetoing events based on a $\Delta R$
matching.
The cross-sections for inclusive $W$- and $Z$-boson production are calculated
with NNLO precision using the \textsc{FEWZ}
program~\cite{Gavin:2010az,Gavin:2012sy} and are estimated to be \SI{12.1}{\nb}
and \SI{1.13}{\nb}, respectively. 
The uncertainty is \SI{4}{\percent}, including the contributions from the PDF
and scale variations.

Samples of diboson ($VV$, $V=W$ or $Z$) events were produced using the
\SHERPA~(v1.4.1) event generator~\cite{Gleisberg:2008ta} with the \ct PDF set,
up to three additional partons in the ME, and a dedicated parton-shower tune
developed by the \SHERPA authors.
The CKKW method~\cite{Hoeche:2009rj} was used to remove overlap between partonic
configurations generated by the ME or the PS.
All three processes are normalised using the inclusive NLO cross-sections
provided by \mcfm~\cite{Campbell:2011bn}, which are \SI{56.8}{\pb} for $WW$,
\SI{7.36}{\pb} for $ZZ$, and \SI{21.5}{\pb} for $WZ$ production. The total
uncertainty for each of the three processes, including scale variations and
uncertainties in the PDF, is estimated to be \SI{5}{\percent}.

\section{Event reconstruction}
\label{sec:object}

In this analysis, \ttbar candidate events are identified by means of isolated
electrons and muons, jets, some of which are possibly $b$-tagged as likely to
contain $b$-hadrons, and sizeable missing transverse momentum.
The definitions of these reconstructed objects, called detector-level objects,
together with the corresponding objects reconstructed using only MC
event generator information, called particle-level objects, are discussed in
this section. The particle-level objects are used to define a fiducial volume.

\subsection{Detector-level object reconstruction}

\textbf{Electrons}: Electron candidates are reconstructed by matching tracks in
the ID to energy deposits (clusters) in the electromagnetic calorimeter
\cite{PERF-2016-01}. Selected electrons are required to satisfy strict quality
requirements in terms of shower shape, track properties and matching quality.
Electron candidates are required to be within $|\eta| < 2.47$, and candidates in
the calorimeter barrel--endcap overlap region, $1.37 < |\eta| < 1.52$, are
excluded. Electrons from heavy-flavour decays, hadronic jets misidentified as
electrons, and photon conversions are the major backgrounds for
high-\pT~electrons associated with a $W$-boson decay. The suppression of these
backgrounds is possible via isolation criteria that require little calorimeter
activity and a small sum of track \pT in an $\eta$--$\phi$ cone around the
electron. The electromagnetic (EM) calorimeter isolation variable is defined as
the scalar sum of the transverse momenta of calorimeter energy deposits within
the cone, corrected by subtracting the estimated contributions from the electron
candidate and from the underlying event and pile-up contributions. The track isolation
variable is defined as the scalar sum of all track transverse momenta
within the cone, excluding the track belonging to the electron
candidate~\cite{PERF-2013-05}.
Thresholds are imposed on the EM calorimeter isolation variable in a cone of
size $\Delta R = 0.2$ around the electron and on the track isolation in a cone
of size $\Delta R = 0.3$.
The isolation requirements imposed on the electron candidates are tuned to
achieve a uniform selection efficiency of \SI{90}{\percent} across electron transverse energy \et and
pseudorapidity $\eta$. The electron pseudorapidity is taken from the associated
track.

\textbf{Muons}: Muon candidates are reconstructed by combining MS tracks
with tracks in the ID, where tracks in the MS and ID are reconstructed
independently~\cite{PERF-2014-05}.
The final candidates are required to be in the pseudorapidity region of
$|\eta|<2.5$.
A set of requirements on the number of hits in the ID must also be
satisfied by muon candidates. An isolation requirement~\cite{Rehermann:2010vq}
is applied to reduce the contribution of muons from heavy-flavour decays. The
isolation variable is defined as the scalar sum of the transverse momenta of all
tracks originating from the primary vertex with \pT above \SI{1}{\gev},
excluding the one matched to the muon, within a cone of size
$\Delta R_{\text{iso}}=\SI{10}{\gev}/\pT(\mu)$, where $\pT(\mu)$ is the
transverse momentum of the muon. Muon candidates are accepted when the
value of the isolation variable divided by the $\pT(\mu)$ is smaller than 0.05.

\textbf{Jets}: Jets are reconstructed using the anti-$k_{t}$
algorithm~\cite{Cacciari:2008gp} implemented in the FastJet package
\cite{Cacciari:2011ma} with a radius parameter of 0.4, using topological
clusters calibrated with the local cell weighting (LCW)
method~\cite{PERF-2014-07} as inputs to the jet finding algorithm. The energies
of the reconstructed jets are calibrated using \pt- and $\eta$-dependent factors
that are derived from MC simulation with a residual in situ calibration based
on data~\cite{PERF-2012-01}. In addition, a pile-up correction is applied to
both the data and Monte Carlo (MC) events to further calibrate the jet before
selection~\cite{PERF-2014-03}.
To reject jets likely to have originated from pile-up, a variable called the jet
vertex fraction ($\mathrm{JVF}$) is defined as the ratio of
$\sum{p_{\mathrm{T},i\in \text{PV}}}$ of all tracks in the jet which originate
from the primary vertex to the $\sum{p_{\mathrm{T},i}}$ of all tracks in the
jet. Only tracks with \pt > \SI{1}{\gev} are considered in the $\mathrm{JVF}$
calculation.
Jets with $|\eta|<2.4$ and $\pT < \SI{50}{\gev}$ are required to have
$|\mathrm{JVF}|> 0.5$.

\textbf{Identification of $b$-quark jets\label{btagging}}:
One of the most important selection criteria for the analysis of events
containing top quarks is the one that identifies jets likely to contain
$b$-hadrons, called $b$-tagging. Identification of $b$-jets is based on the
long lifetime of $b$-hadrons, which results in a significant flight path
length and leads to reconstructable secondary vertices and tracks with large
impact parameters relative to the primary vertex.
In this analysis, a neural-network-based algorithm is used at a working point
corresponding to a $b$-tagging efficiency in the simulated \ttbar events of
\SI{70}{\percent}, a $c$-jet rejection factor of \num{5}
and light-flavour jet rejection factor of \num{140}~\cite{PERF-2012-04}.

\textbf{Missing transverse momentum}: The missing transverse momentum is a
measure of the momentum of the escaping neutrinos.
It also includes energy losses due to detector inefficiencies, leading to the
mismeasurement of the true transverse energy $\et$ of the detected final-state
objects. The missing transverse momentum vector, \metvec, is calculated as the
negative vector sum of the transverse momenta of reconstructed and calibrated
physics objects, i.e.
electrons, muons and jets as well as energy deposits in the calorimeter which
are not associated with physics objects~\cite{PERF-2014-04}.
The magnitude of the missing transverse momentum vector is defined as $\met = |
\metvec |$.

A procedure to remove overlaps between physics objects is applied, where jets
overlapping with identified electron candidates within a cone of size $\Delta R
= 0.2$ are removed from the list of jets, as the jet and the electron are very
likely to correspond to the same physics object. In order to ensure the
selection of isolated charged leptons, further overlap removals are
applied.
If electrons are still present with distance $\Delta R < 0.4$ to a jet,
they are removed from the event. Muons overlapping with a jet within $\Delta R < 0.4$ are discarded from the event.

\subsection{Particle-level object reconstruction}
\label{sec:fid}
Particle-level objects are defined using stable particles with a mean lifetime
greater than \SI[exponent-product=\cdot]{0.3e-10}{\second}.
Selected leptons are defined as electrons, muons or neutrinos originating from
the $W$-boson decay, including those that originate from a subsequent
$\tau$-lepton decay.
Leptons from hadron decays either directly or via a hadronic $\tau$ decay are
excluded.
The selected charged lepton is combined with photons within $\Delta R <
0.1$, which implies that the final four-momentum vector is the vector sum of the
associated photons and the original lepton four-vector.
Finally the charged lepton is required to have $\pT > \SI{25}{\GeV}$ and $|\eta|
< 2.5$.

Particle-level jets are reconstructed using the anti-$k_t$
algorithm with a radius parameter of $R=0.4$. All stable
particles are used for jet clustering, except the selected leptons
(electrons, muons or neutrinos) and the photons associated with the charged leptons.
This implies that the energy of the particle level $b$-jet is close to
that of the $b$-quark before hadronisation and fragmentation.
Each jet is required to have $\pT > \SI{25}{\GeV}$ and $|\eta| < 2.5$.
             
Events are rejected if a selected lepton is at a distance $\Delta R < 0.4$
to a selected jet.

The fiducial volume is defined by selecting exactly one electron or muon and at
least three particle-level jets. Setting the minimum number of
particle-level jets to three minimises the extrapolation uncertainty
going from the detector-level volume to the particle-level fiducial volume.
In this case the fraction of events which are selected in the
detector-level volume and not selected in the particle-level fiducial volume
is of the order of \SI{10}{\percent}.

\section{Event selection and classification \label{sec:sel}}

This section describes the selection of \ttbar candidate events.
The datasets used in this analysis are obtained from single-electron or
single-muon triggers. For the electron channel, a calorimeter energy cluster
needs to be matched to a track, and the trigger electron candidate is required
to have $\ET > \SI{60}{\GeV}$ or $\ET > \SI{24}{\GeV}$ with additional isolation requirements~\cite{PERF-2011-02}.
The single-muon trigger~\cite{TRIG-2012-03} requires either an isolated muon
with $\pT > \SI{24}{\gev}$ or a muon with $\pT > \SI{36}{\gev}$.

Each event is required to have at least one vertex reconstructed from at least
five tracks, where the \pt of each track is above \SI{400}{\mev}.
The vertex with the largest sum of \pTsq of the associated tracks is chosen as the primary vertex. 
Events containing any jets with $\pT > \SI{20}{\GeV}$ failing to satisfy quality criteria
defined in Ref.~\cite{PERF-2011-03} are rejected, in order to suppress
background from beam--gas and beam-halo interactions, cosmic rays and
calorimeter noise.

In order to select \ttbar events in the lepton+jets channel, exactly one
electron or muon with $\pT > \SI{25}{\gev}$ is required. In addition
to the requirements explained in Sect.~\ref{sec:object}, the $\Delta R$ value
between the reconstructed lepton and the trigger-lepton has to be smaller than \num{0.15}.
Events containing an electron candidate and a muon candidate sharing an ID track
are discarded.

Furthermore, events must have at least four jets with $\pT > \SI{25}{\gev}$
and $|\eta|<2.5$.
At least one of the jets has to be $b$-tagged. To enhance the fraction of events
with a leptonically decaying $W$ boson, events are required to have $\MET >
\SI{25}{\gev}$ and the transverse mass \mtw of the lepton--$\MET$ pair is
required to be
\begin{equation*}
  \mtw = \sqrt{2 \pT(\ell) \cdot \MET 
  \left[1-\cos \left( \Delta \phi \left(\vec{\ell}, \metvec \right) \right)
  \right]}\, \ > \ \SI{30}{\gev},
\label{eq:mTW}
\end{equation*} 
where $\pT(\ell)$ is the transverse momentum of the charged lepton and $\Delta
\phi $ is the angle in the transverse plane between the charged lepton and the
\metvec vector.

The measurement of the \ttbar cross-section is performed by splitting the
selected sample into three disjoint signal regions. These have different
sensitivities to the various backgrounds, to the production of additional
radiation, and to detector effects.

\begin{itemize}
\item \textbf{\SRone:} $\ge4$ jets, 1 $b$-tag \\
In this region, 
events with at least four jets of which exactly one is $b$-tagged are
selected.
This region has the highest background fraction of all three signal regions,
with \Wjets being the dominant background. This signal region has the highest
number of selected events.
                        
\item \textbf{\SRtwo:} 4 jets, 2 $b$-tags \\
In this region, events with exactly four jets of which exactly two are
$b$-tagged are selected. The background is expected to be small in this region
and this allows an unambiguous matching of the reconstructed objects to the
top-quark decay products.
In particular, the two untagged jets are likely to originate from the
hadronically decaying $W$ boson. The reconstructed $W$-boson mass is sensitive
to the jet energy scale and to additional radiation.

\item \textbf{\SRthr:} $\ge4$ jets, $\ge 2$ $b$-tags (excluding events from
\SRtwoNoSpace)\\
In the third region, events are required to have at least four jets with  
at least two $b$-tagged jets. Events with exactly four jets and two
$b$-tags are assigned to \SRtwoNoSpace. This region includes \ttbar events with
extra gluon radiation, including \ttbar + heavy-flavour production, and is
sensitive to the efficiency of misidentifying $c$-jets, originating mainly from the
$W\rightarrow cs$ decay, as $b$-jets. The expected background is the smallest
among the signal regions.

\end{itemize}
For the determination of the \ttbar cross-section 
a discriminating variable in each signal region is defined, as explained in
Sect.~\ref{sec:nn}.
The number of \ttbar events is extracted using a simultaneous fit of three
discriminating-variable distributions, one from each signal region, to data. In
order to reduce systematic uncertainties due to the jet energy scale and $b$-tagging efficiency, their
effects on the signal and background distributions are parameterised with
nuisance parameters, which are included in the fit.

\section{Background modelling and estimation}
\label{sec:nn_background}

The dominant background to \ttbar production in the lepton+jets final state is
\Wjets production.
This analysis uses a sample defined from data to model the shapes of
the discriminating-variable distributions for this background, while the
normalisation in each signal region is determined in the final fit. The multijet
background process, which is difficult to model in the simulation, is also
modelled using data and normalised using control regions. All
remaining backgrounds are determined using simulated events and theoretical
predictions.

The method used to model the \Wjets background shape
from data is based on the similarity of the production and decay of the $Z$ boson
to those of the $W$ boson.

First, an almost background-free
\Zjets sample is selected in the following way:
\begin{itemize}
    \item Events are required to contain exactly two oppositely charged leptons
    of the same flavour, i.e. \ee or \mumu, and
    \item the dilepton invariant mass $m(\ell\ell)$ has to be consistent with
    the $Z$-boson mass ($80 < m(\ell\ell) < \SI{102}{\gev}$).
\end{itemize}
These events are then \enquote{converted} into \Wjets events.
This is achieved by boosting the leptons of the $Z$-boson decay
into the $Z$ boson's rest frame, scaling their momenta to that of a lepton decay
from a $W$ boson by the ratio of the boson masses and boosting the leptons back into
the laboratory system. The scaled lepton momenta are given by
\begin{equation*}
\vec{p'^{\ast}}_{\ell_i}=\frac{m_W}{m_Z} \vec{p^{\ast}}_{\ell_i},
\end{equation*}
where $\vec{p^{\ast}}_{\ell_i}$ is the momentum vector of lepton $i$ in the
$Z$-boson's rest frame, $m_W$ and $m_Z$ are the masses of the $W$- and
$Z$-bosons respectively, and $\vec{p'^{\ast}}_{\ell_i}$ is the scaled momentum
vector of lepton $i$ in the $Z$-boson's rest frame.

After this conversion, one of the leptons is randomly chosen to be removed, and
the \metvec vector is recalculated.
Finally, the event selection requirements discussed in Sect.~\ref{sec:sel} are
applied, except for the $b$-tagging requirement.
In the following, this sample is referred to as the \ZtoW sample.

\begin{figure}[htbp]
	\centering
	\subfigure[]{
	  \includegraphics[width=0.45\textwidth]{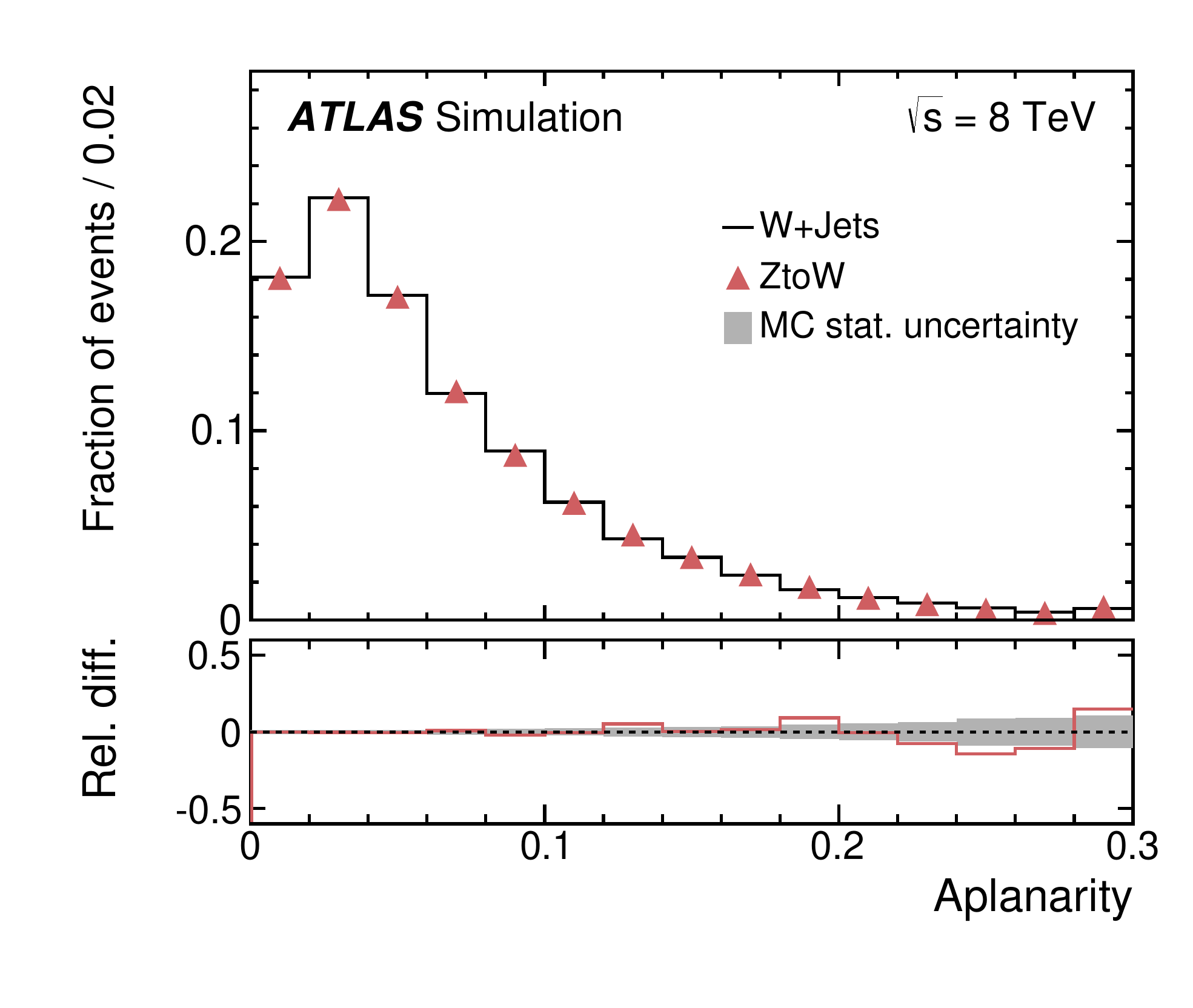}
	  \label{subfig:ZtoWVal_aplanarity} 
	}
	\subfigure[]{
    \includegraphics[width=0.45\textwidth]{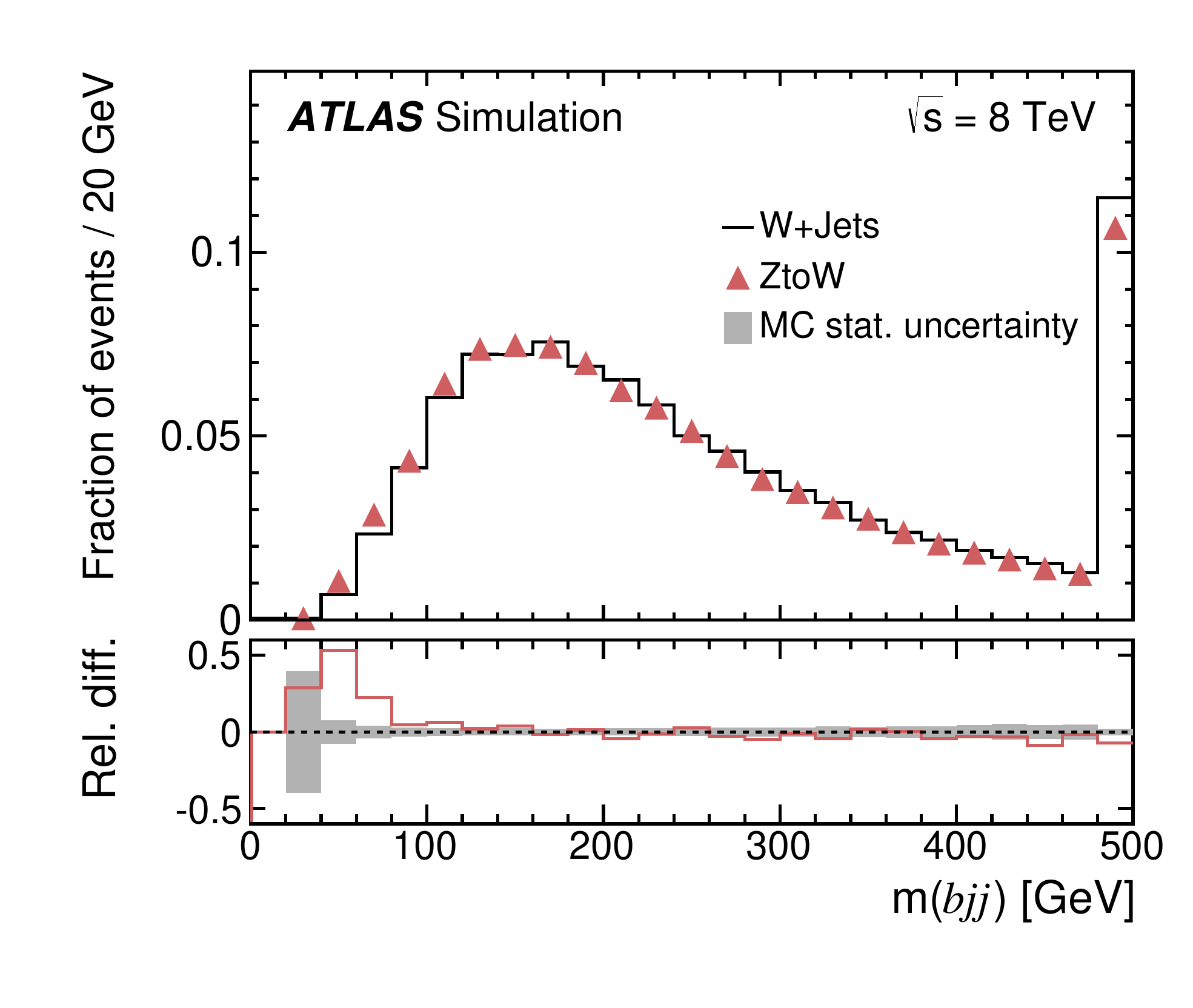}
    \label{subfig:ZtoWVal_hadtopmass} 
    }
    \vspace{-0.5cm}
\caption{Probability densities of \subref{subfig:ZtoWVal_aplanarity} the
aplanarity and \subref{subfig:ZtoWVal_hadtopmass} the mass distribution of the
hadronically decaying top-quark candidates for simulated \Wjets events with at
least four jets and at least one $b$-tag and \ZtoW events derived from a
simulated \Zjets sample with at least four jets and no $b$-tagging
requirement.
The lower histogram shows the relative difference between the numbers of \ZtoW
and \Wjets events in each bin with respect to the number of \Wjets
events.
The grey error band represents the Monte Carlo statistical uncertainty of the
\Wjets sample.
Events beyond the $x$-axis range are included in the last bin.}
	\label{fig:ZtoWValidation}
\end{figure}

Detailed studies in simulation and in validation regions are performed. As an
example, two important variables, discriminating between \Wjets and \ttbar
events, are compared between simulated \Wjets events with at least four jets and
at least one $b$-tag and \ZtoW events derived from a simulated \Zjets sample
with at least four jets and no $b$-tagging requirement.
Distributions of these variables are shown in Fig.~\ref{fig:ZtoWValidation}:
the aplanarity event-shape variable and the mass of the hadronically decaying top-quark
candidate.
Details about the top-quark reconstruction are given in Sect.~\ref{sec:nn}.
The aplanarity is defined as 
\begin{eqnarray}
\label{eq:apl}
          A & = & \frac{3}{2}\lambda_3 ,
\end{eqnarray}  
where $\lambda_3$ is the smallest eigenvalue of the sphericity tensor,
defined by 
\begin{eqnarray*}
S^{\alpha \beta } & = & \frac{\sum_i p_i^{\alpha } p_i^{\beta} }{\sum_i\mid p_i
\mid^2 }.
\end{eqnarray*}         
Here, $\alpha$ and $\beta$ correspond to the $x$, $y$ and $z$ momentum
components of final-state object $i$ in the event, i.e. the
jets, the charged lepton and the reconstructed neutrino (see Sect. \ref{sec:nn}).

Residual differences between the shapes of the \Wjets and \ZtoW templates are
accounted for as a systematic uncertainty in the analysis.
Since the method only provides shape information, the expected number of events
for the \Wjets process in the signal regions is obtained from
the acceptance of simulated samples using \ALPGEN + \PYTHIA 
and normalised using the inclusive NNLO \Wjets
cross-section as described in
Sect.~\ref{sec:samples}. 
These numbers are used to define the nominal background 
yield prior to the fit and used as initial values for the fit 
in the final statistical analysis.

Multijet events may be selected if a jet is misidentified as an isolated lepton or 
if the event has a non-prompt lepton that appears to be isolated 
(these two sources of background are referred to as fake leptons).
The normalisation of the multijet background is obtained from a fit to the
observed \MET distribution in the electron channel or to the \mtw~distribution
in the muon channel in the signal regions.
In order to construct a sample of multijet background events, 
different methods are adopted for the electron and muon channels.

The \enquote{jet-lepton} method~\cite{ATLAS-CONF-2014-058} is used to model the
background due to fake electrons using a dijet sample simulated with the
\PYTHIAV{8} event generator~\cite{Sjostrand:2014zea}.
A jet that resembles the electron has to have $\et > \SI{25}{\gev}$ and be
located in the same $\eta$ region as the signal electrons. The jet
energy must have an electromagnetic fraction of between \num{0.8} and
\num{0.95}.
The event is accepted if exactly one such jet is found, and if the event passes
all other selection requirements as described above, except the one on \met. The
yield of the multijet background in the electron-triggered data sample is then
estimated using a binned maximum-likelihood fit to the \met distribution using
the template determined from the selected events in the dijet simulated sample.
In order to improve the modelling of the $\eta(\ell)$ distribution of the
\enquote{jet-lepton} model in \SRoneNoSpace, the fit is done separately in the
barrel region ($|\eta| \leq 1.37$) and in the endcap region ($|\eta| >
1.52$).
The fits for \SRtwo and \SRthr are performed inclusively in $|\eta|$ due to the lower
number of selected events.

The \enquote{anti-muon} method~\cite{ATLAS-CONF-2014-058} uses a dedicated
selection on data to enrich a sample in events that contain fake muons in order
to build a multijet model for muon-triggered events.
Some of the muon identification requirements differ from those for signal muon
candidates. The calorimeter isolation requirement is inverted, while keeping the
total energy loss of the muon in the calorimeters below $\SI{6}{\gev}$, and the
requirement on the impact parameter is omitted. The additional application of
all other event selection requirements mentioned in Sect.~\ref{sec:sel} results
in a sample that is highly enriched in fake muons from multijet events, but contains only a small
fraction of prompt muons from $Z$- and $W$-boson decays.
The yield of the multijet background in the muon-triggered data sample is
estimated from a maximum-likelihood fit to the \mtw distribution using
the template determined from the selected multijet events in the data sample.
A different fit observable (\mtw) in the muon channel is used, since it provides a better
modelling of the multijet background than the \met observable used in the
electron channel.

\begin{figure}[htbp]
	\centering
	\subfigure[]{
	  \includegraphics[width=0.45\textwidth]{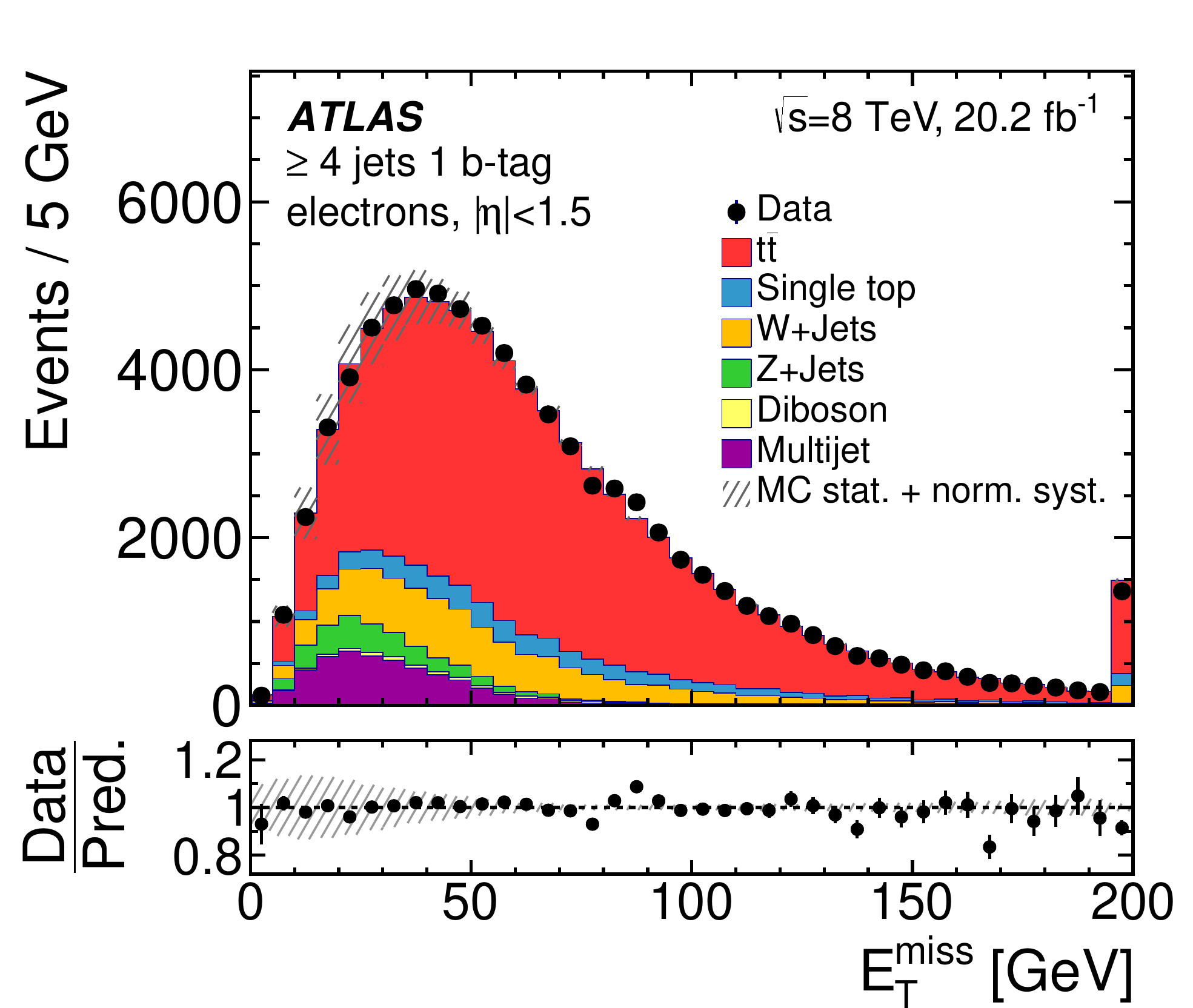}
	  \label{subfig:el_ch1_qcdfit_lowEta} 
	}
	\subfigure[]{
    \includegraphics[width=0.45\textwidth]{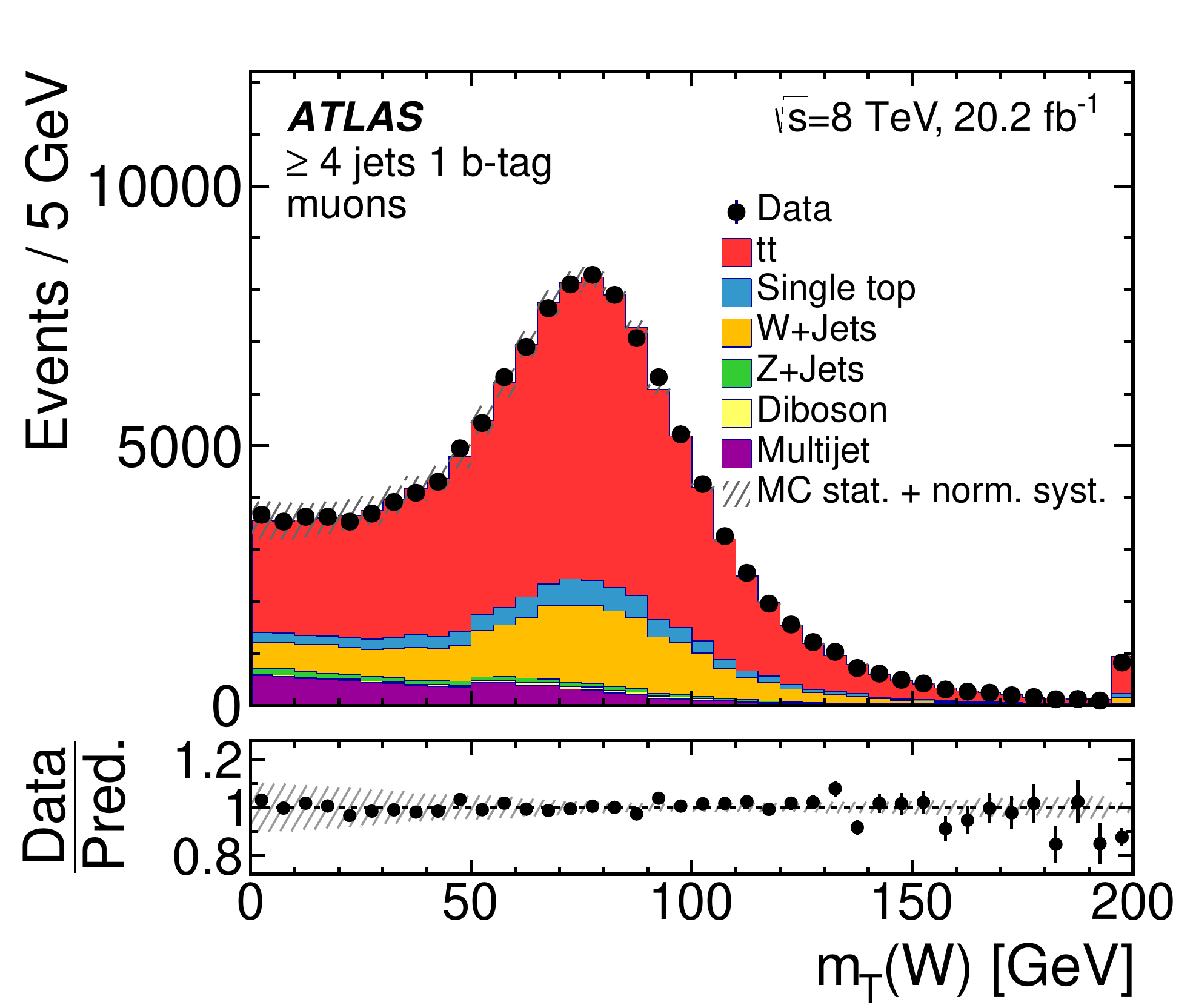}
    \label{subfig:muo_ch1_qcdfit} 
    }
	\subfigure[]{
	\includegraphics[width=0.45\textwidth]{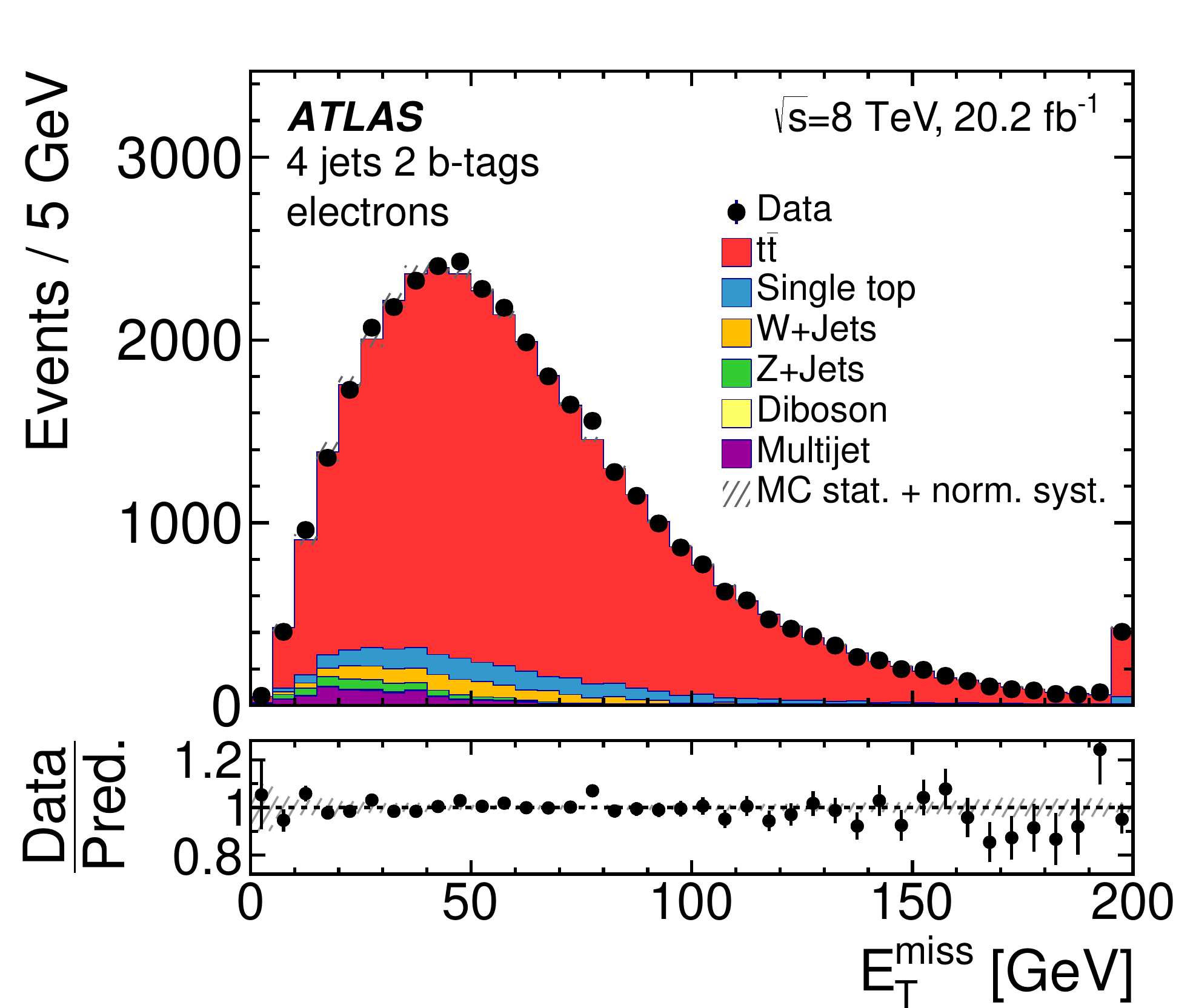}
	     \label{subfig:el_ch2_qcdfit}
	}
	\subfigure[]{
    \includegraphics[width=0.45\textwidth]{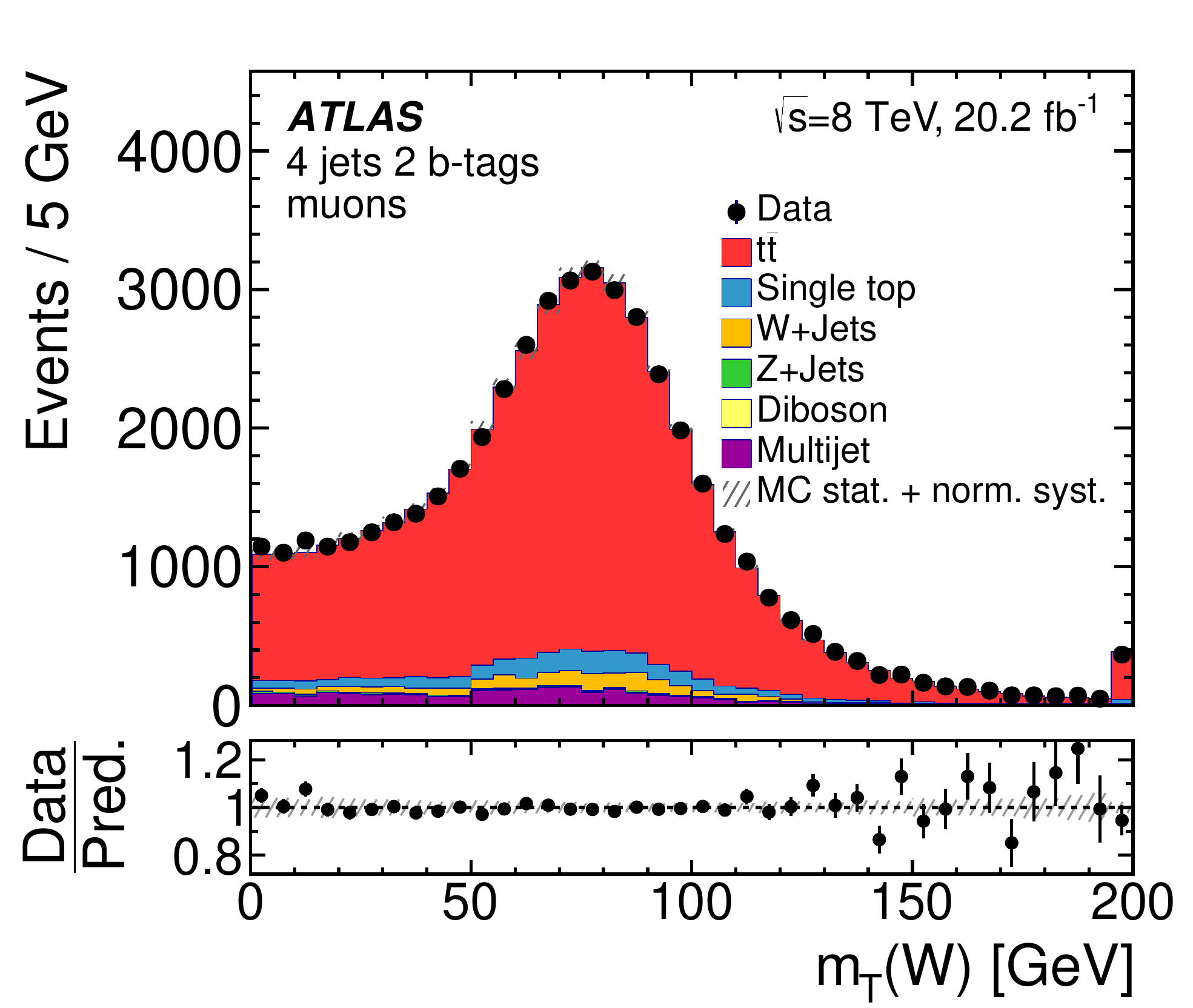}
     \label{subfig:muo_ch2_qcdfit}
    }
	\subfigure[]{
	\includegraphics[width=0.45\textwidth]{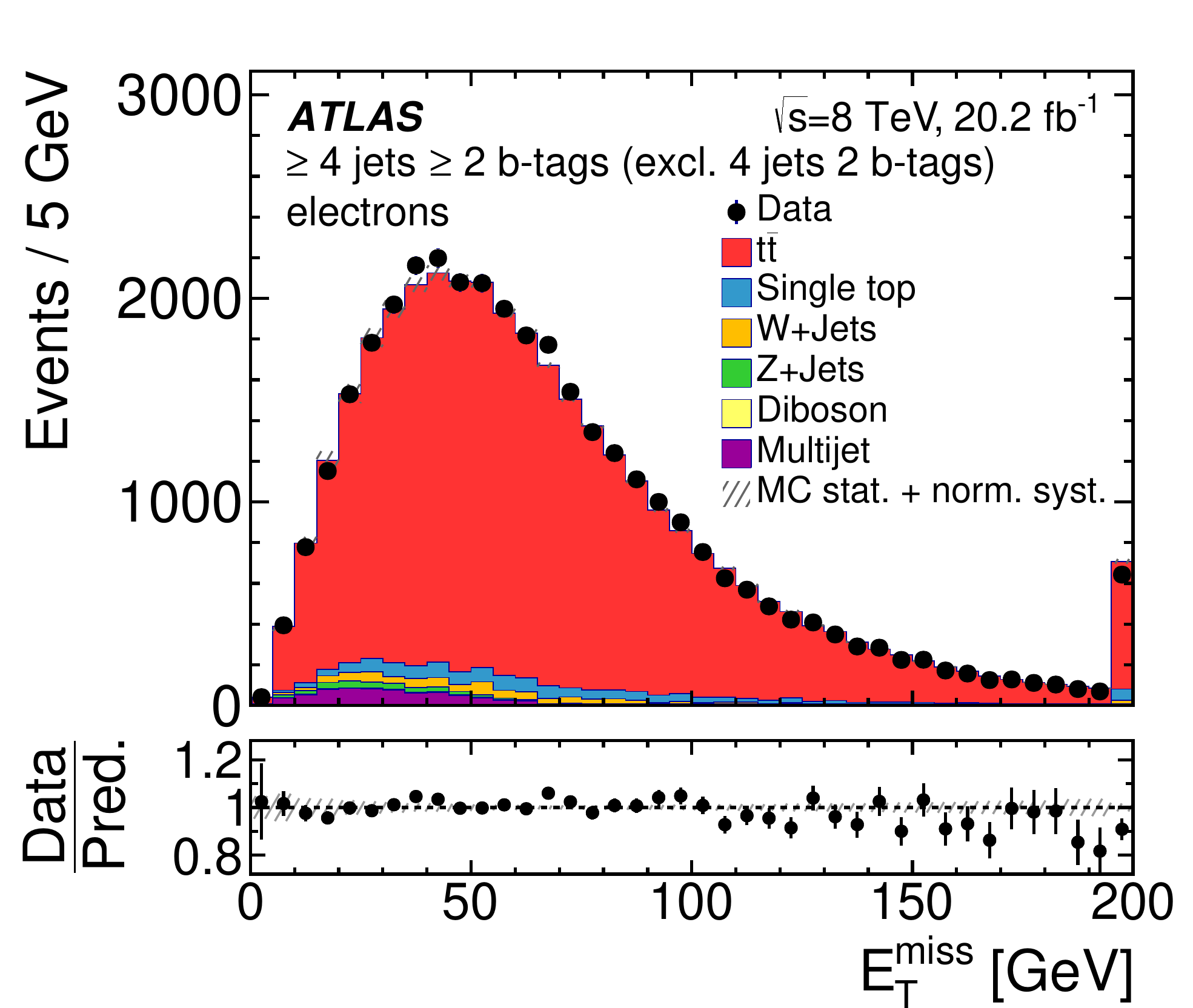}
	     \label{subfig:el_ch3_qcdfit}  
	}
    \subfigure[]{
    \includegraphics[width=0.45\textwidth]{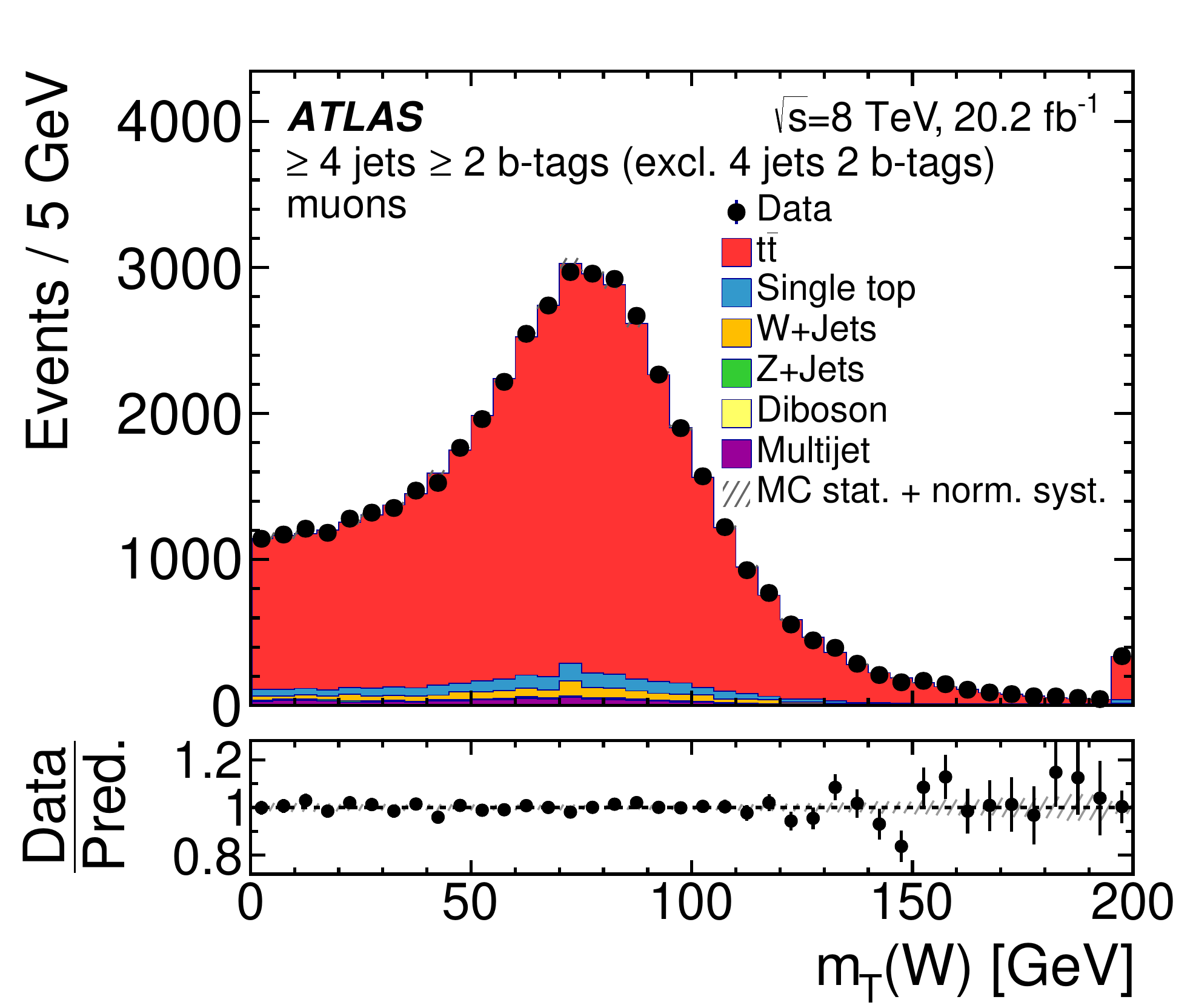}
         \label{subfig:muo_ch3_qcdfit}  
    }
    \vspace{-0.5cm}
	\caption{
	Observed and simulated (left) \met distributions in the electron channel and
	(right) \mtw distributions in the muon channel, normalised to the result of the
	binned maximum-likelihood fit, \subref{subfig:el_ch1_qcdfit_lowEta} for the
	barrel region in \SRoneNoSpace, \subref{subfig:muo_ch1_qcdfit} in
	\SRoneNoSpace,
	\subref{subfig:el_ch2_qcdfit}-\subref{subfig:muo_ch2_qcdfit} in \SRtwoNoSpace,
	and \subref{subfig:el_ch3_qcdfit}-\subref{subfig:muo_ch3_qcdfit} in \SRthrNoSpace.
    The hatched error bands represent the uncertainty due to the
    sample size and the normalisation of the backgrounds.
    The ratio of observed to predicted (Pred.) number of events in
    each bin is shown in the lower histogram.
    Events beyond the $x$-axis range are included in the last bin.}
	\label{fig:multijet_fits}
\end{figure}

In both methods to obtain the multijet background normalisation, the
multijet template is fitted together with templates derived from MC simulation for the
\ttbar and \Wjets processes. The \ttbar and \Wjets rate uncertainties,
obtained from theoretical cross-section uncertainties, are accounted for in the fitting process in the form of
constrained normalisation factors.
The rates for \Zjets, single-top-quark processes, and $VV$ processes are fixed
to the predictions as described in Sect.~{\ref{sec:samples}}.
For the fits in \SRtwo and \SRthrNoSpace, the \Wjets process is fixed as well,
since the predicted yield is very small in these signal regions.
The resulting fitted rate of \ttbar events is in agreement within the
statistical uncertainty with the result of the final estimation of the \ttbar cross-section
and therefore does not bias the result. Distributions of the fitted observable, normalised to the fit
results, are shown in Fig.~\ref{fig:multijet_fits}.

The \enquote{matrix} method~\cite{ATLAS-CONF-2014-058} is used as an alternative
method to estimate systematic uncertainties in the multijet background estimate.
It provides template distributions and estimates of the number of multijet
events in \SRoneNoSpace. Differences between the fitting method and the
\enquote{matrix} method are taken into account as systematic uncertainties
yielding a normalisation uncertainty of \SI{67}{\percent}.
Due to the small number of events for the \enquote{matrix} method in \SRtwo and
\SRthr an uncertainty of \SI{50}{\percent} is assigned, based on comparisons of
the rates obtained using alternative methods described in previous
analyses~\cite{ATLAS-CONF-2014-058}.

As a result of the above procedure, the fraction of the total background
estimated to originate from multijet events for $\met > \SI{25}{\GeV}$ and $\mtw
> \SI{30}{\GeV}$ is $(\num{5.4}\pm\num{3.0})\si{\percent}$ in \SRoneNoSpace,
$(\num{2.6}\pm\num{1.3})\si{\percent}$ in \SRtwo and
$(\num{1.5}\pm\num{0.8})\si{\percent}$ in \SRthrNoSpace. All other processes,
namely \ttbar and single top-quark production, \Zjets and $VV$ production, are
modelled using simulation samples as described in Sect. \ref{sec:samples}.

Table~\ref{tab:event-composition} summarises the event yields in the
three signal regions for the \ttbar signal process and each of the background processes.
The yields, apart from the multijet background, are calculated using the
acceptance from MC samples normalised using their respective theoretical
cross-sections as discussed in Sect.~\ref{sec:samples}. The quoted uncertainties
correspond to the statistical uncertainties in the Monte Carlo samples, except
in the case of the multijet background where they correspond to the
uncertainties in the background estimate.
\begin{table}[h]
\sisetup{round-mode=figures, retain-explicit-plus}
  \caption{Event yields for the three signal regions. The multijet background
  and its uncertainty are estimated from the \met or \mtw fit to data.
  All the other expectations 
  are derived using theoretical cross-sections, and the corresponding
  Monte Carlo sample statistical uncertainty.}
\label{tab:event-composition}
\begin{center}
\begin{tabular}{l
r@{$\,\pm\,$}S[table-format=4.0]
r@{$\,\pm\,$}S[table-format=4.0]
r@{$\,\pm\,$}S[table-format=4.0]
}
\toprule
Process &  \multicolumn{2}{c}{\SRone} & 
 \multicolumn{2}{c}{\SRtwo} &
 \multicolumn{2}{c}{\SRthr} \\
\midrule
\ttbar & {\numRF{133314}{5}} & 365 & {\numRF{63062}{4}} & 251 &
{\numRF{59306}{4}} & 244 \\
Single top & {\numRF{11020}{4}} & 105 & 3728 & 61 & 2593 & 51 \\
\Wjets & {\numRF{29872}{4}} & 173 & 2382 & 49 & 1592 & 40 \\
\Zjets & 3569 & 60 & 406 & 20 & 270 & 16 \\
Diboson & 1339 & 37 & 135 & 12 & 112 & 11 \\
Multijet & {\numRF{10256}{3}} & 6872 & {\numRF{1943}{3}} & 972 & {\numRF{1051}{3}} & 526
\\
\midrule
Total expected & {\numRF{189370}{4}} & 6885 & {\numRF{71657}{3}} & 1007 &
{\numRF{64924}{4}} & 583 \\
\midrule
Observed & \multicolumn{1}{S[table-format=6.0,
 round-mode=off]@{\phantom{$\,\pm\,$}}}{192686}  & &
 \multicolumn{1}{S[table-format=5.0,
 round-mode=off]@{\phantom{$\,\pm\,$}}}{72978} & &
 \multicolumn{1}{S[table-format=5.0,
 round-mode=off]@{\phantom{$\,\pm\,$}}}{70120} \\
\bottomrule
\end{tabular}
\end{center}
\end{table}

\section{Discriminating observables}
\label{sec:nn}

In order to further separate
the signal events from background events in \SRone and \SRthrNoSpace, the output
distribution of an artificial neural network (NN)~\cite{Feindt:2004wla,Feindt:2006pm} is used. 
A large number of potential NN
input variables are studied for their discriminating power between \Wjets and
\ttbar and the compatibility of their distributions between simulated \Wjets
events with at least one $b$-tag and \ZtoW events with no $b$-tagging
requirement. The observables investigated are based on invariant masses of jets
and leptons, event shape observables and properties of the reconstructed top quarks. 

In \SRone and \SRthrNoSpace, the semileptonically decaying top quark is
reconstructed. First, the leptonically decaying $W$ boson's four-momentum is
reconstructed from the identified charged lepton's four-momentum and the \met
vector, the latter representing the transverse momentum of the neutrino.
The unmeasured $z$-component of the neutrino momentum $p_z(\nu)$ is inferred by
imposing a $W$-boson mass constraint on the lepton--neutrino system, leading to
a two-fold ambiguity. In the case of two real solutions, the one with the lower
$|p_z(\nu)|$ is chosen.
In the case of complex solutions, which can occur due to the finite \met
resolution, a fit is performed that rescales the neutrino $p_x$ and $p_y$ such
that the imaginary radical vanishes, at the same time keeping the transverse
components of the neutrino's momentum as close as possible to the $x$- and
$y$-components of \met. To reconstruct the semileptonically decaying top quark,
the four jets with the highest \pt are selected and the one with the smallest
$\Delta R$ to the charged lepton is chosen to be the $b$-jet. The
semileptonically decaying top quark is then reconstructed by adding the
four-momentum of the $W$ boson and the chosen $b$-jet.
The hadronically decaying top quark is reconstructed by adding the four-momenta
of the remaining three highest-\pT jets.

\begin{table}[t]
\begin{center}
\caption{\label{tab:trainingVars} List of the seven input variables of the NN,
ordered by their discriminating power.}
\begin{tabular}{ll}
\hline
Variable & Definition \\
\hline
$m_{12}$ &  The smallest invariant mass between jet pairs. \\
$\cos(\theta^*)_{bjj}$ & Cosine of the angle between the hadronic top-quark
momentum and the beam direction\\
   &  in the \ttbar rest frame. \\
$m(\ell \nu b)$ & Mass of the reconstructed semileptonically decaying top quark. \\
$A$ & Aplanarity, as defined in Eq.~(\ref{eq:apl}) \\
$m(bjj)$ &  Mass of the reconstructed hadronically decaying top quark. \\
$m_{\ell1}$ &  The smallest invariant mass between the charged lepton and a
jet.\\
$m_{23}$ &  The second  smallest invariant mass between jet pairs. \\
\hline
\end{tabular}
\end{center}
\end{table}

Seven observables are finally chosen as input variables to the NN (see
Table~\ref{tab:trainingVars}). The choice was made by studying the correlations
between the potential input variables and choosing the ones with small
correlations that still provide a good separation between the signal and the
background events. The NN infrastructure consists of one input node for each
input variable plus one bias node, eight nodes in the hidden layer, and one
output node, which gives a continuous output $o_{\mathrm{NN}}$ in the interval [$0,1$].
For the training of the NN, an equal number of simulated \ttbar events and \ZtoW
events are used. The training is performed in an inclusive phase space with $\ge
4$ jets and $\ge 1$ $b$-tag to cover the whole phase space and achieve an
optimal separation power in both signal regions.

\begin{figure}[t]
	\centering
	\subfigure[]{
	\includegraphics[width=0.45\textwidth]{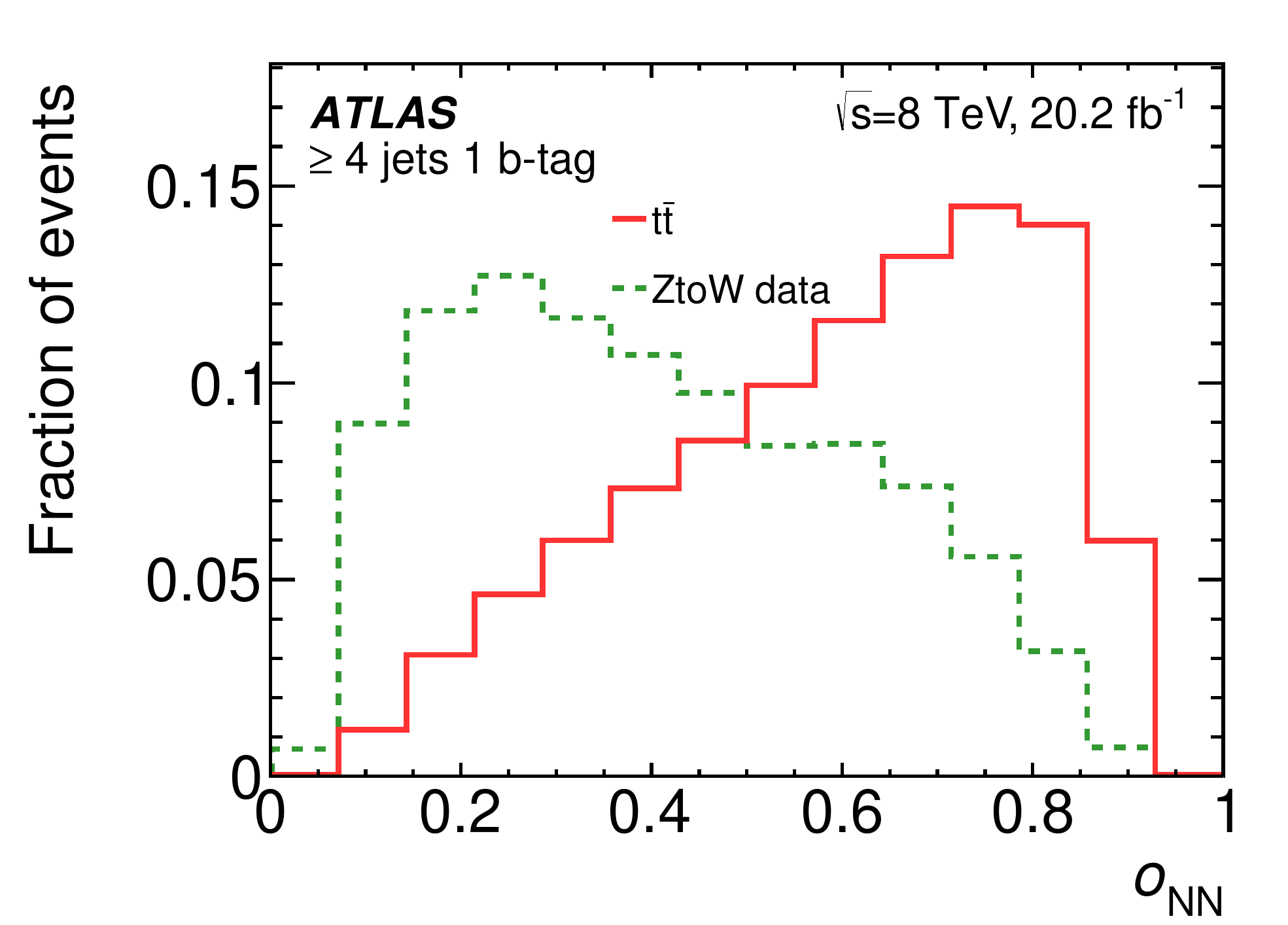}
	 \label{subfig:NN_shape_ch1} 
	}
    \subfigure[]{
    \includegraphics[width=0.45\textwidth]{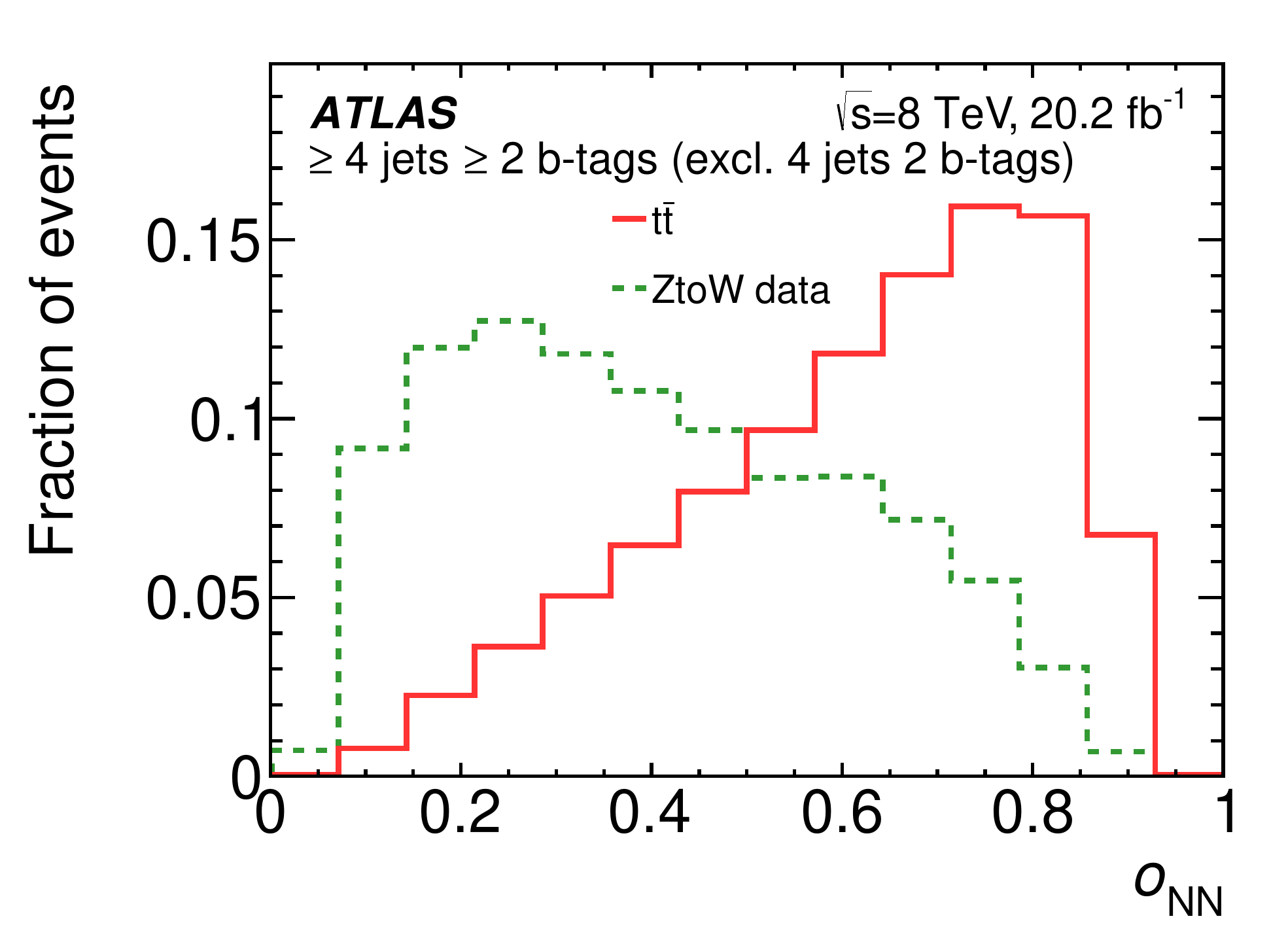} 
    \label{subfig:NN_shape_ch3}
    }
	\caption{ 
	Probability densities of the neural-network discriminant
	$o_{\mathrm{NN}}$ for the simulated \ttbar signal process and the \Wjets background
	process derived from data using converted \Zjets events
	\subref{subfig:NN_shape_ch1} for \SRone and \subref{subfig:NN_shape_ch3} for \SRthrNoSpace.}
	\label{fig:NN_shape}
\end{figure}

The discriminating power of
the NN between \ZtoW and \ttbar events can be seen in Fig.~\ref{fig:NN_shape}
for \SRone and \SRthrNoSpace.

\begin{figure}[t]
    \centering
    \subfigure[]{
    \includegraphics[width=0.45\textwidth]{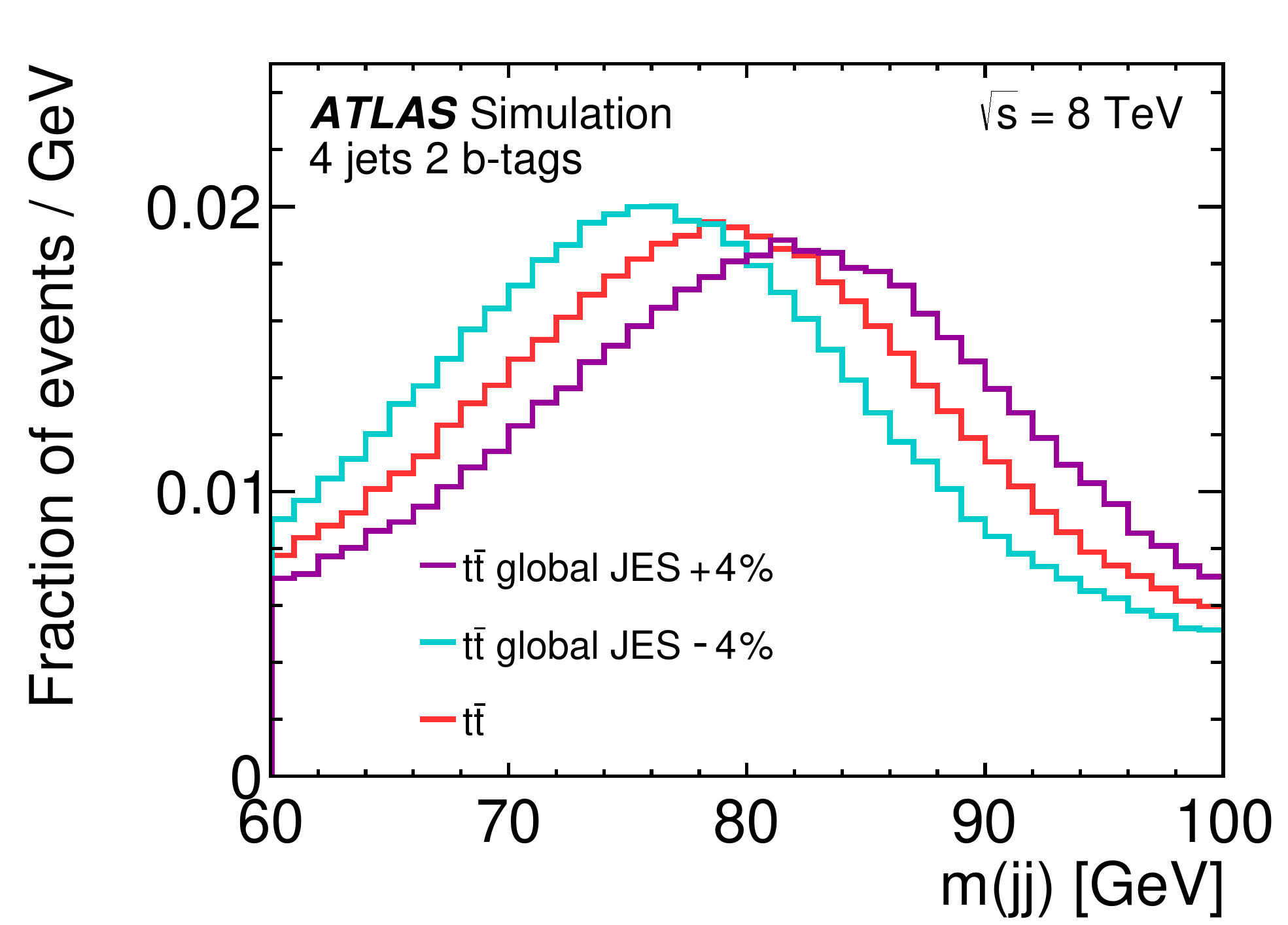}
    \label{subfig:mjj_jes} 
    }
    \subfigure[]{
    \includegraphics[width=0.45\textwidth]{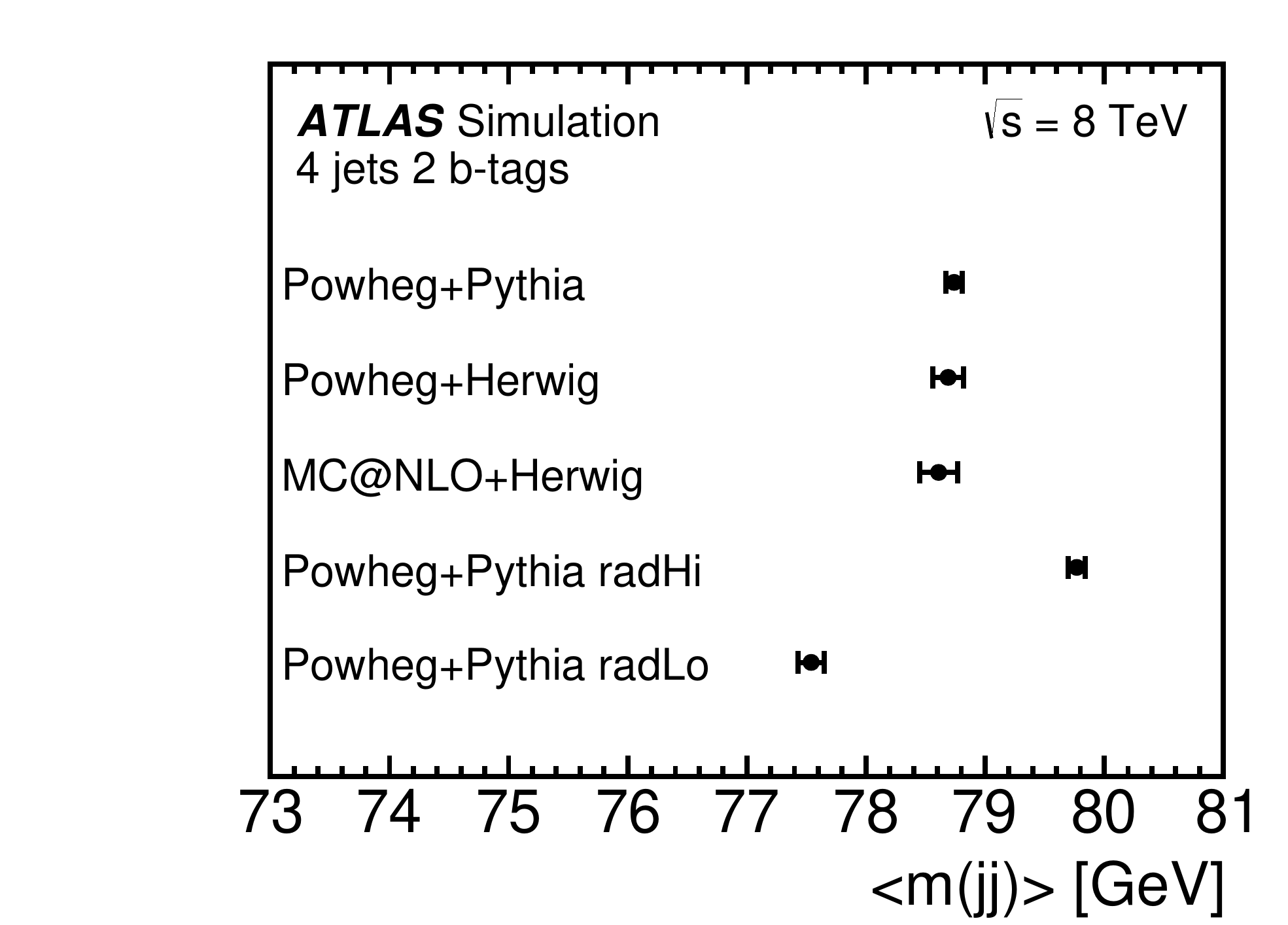}
    \label{subfig:mjj_gen}
    } 
    \caption{ 
    \subref{subfig:mjj_jes} Probability densities of the $m(jj)$ distribution
    from the \ttbar signal process for three different values of the JES,
    where events beyond the $x$-axis range are not shown and the range is restricted to show
    the peak.
    \subref{subfig:mjj_gen} Mean value of the fit to the $m(jj)$ distribution
    using a Gaussian distribution for different signal generator set-ups. The
    uncertainties shown are statistical only.}
\end{figure}

In \SRtwo, a different distribution is used as discriminant in
the final fit. Inspired by measurements of the top-quark mass, 
where the invariant mass of the two untagged
jets $m(jj)$ is frequently utilised to reduce the impact of the jet energy scale
(JES)
uncertainty~\cite{TOPQ-2013-02,CMS-TOP-14-022,Abazov:2008ds,Aaltonen:2012va},
this approach is also followed here.
The dependency of $m(jj)$ on the JES is shown in Fig.~\ref{subfig:mjj_jes} using
simulated \ttbar events with modified global JES correction factors. Here the
energy of the jets is scaled by $\pm \SI{4}{\percent}$. Additionally,
the mean of the $m(jj)$ distribution is sensitive to the amount of additional
radiation.
A comparison of the mean value of a Gaussian distribution fitted to the $m(jj)$
distribution in the range of $\SI{60}{\gev} < m(jj) < \SI{100}{\gev}$ for
different generator set-ups is presented in Fig.~\ref{subfig:mjj_gen}.
It can be seen that the mean value is compatible for different generator
set-ups, but varies for different settings of the parameters controlling the
initial- and final-state radiation.
For these reasons, the $m(jj)$ observable is used as the discriminant in
\SRtwoNoSpace.

Finally, the ratio of single to double $b$-tagged events, i.e. the ratio of
events in \SRone to the sum of events in \SRtwo and \SRthrNoSpace, is sensitive
to the $b$-tagging efficiency. The effect of the $b$-tagging efficiency is
parameterised with a nuisance parameter in the final fit. Since only two
$b$-jets are present in \ttbar events, any additional $b$-tagged jets
originate either from heavy-flavour production in the parton shower or from
mistagged $c$-hadrons.
The inclusion of events with more than two $b$-tags in \SRthr gives a small sensitivity
to heavy-flavour production in the parton shower.

\section{Sources and estimation of systematic uncertainties}
\label{sect:sys}

Several sources of systematic uncertainty affect the \ttbar cross-section
measurement.
In addition to the luminosity determination, they are related to the modelling
of the physics objects, the modelling of \ttbar production and the understanding of the
background processes. All of them affect the yields and kinematic
distributions (shape of the distributions) in the three signal regions.
 
\subsection{Physics objects modelling}
Systematic uncertainties associated with reconstructed jets, electrons and
muons, due to residual differences between data and MC simulations
after calibration, and uncertainties in corrective scale factors are propagated
through the entire analysis.

Uncertainties due to the lepton trigger, reconstruction and selection
efficiencies in simulation are estimated from measurements of the efficiency in data using 
$Z\rightarrow\ell\ell$ decays. 
The same processes are used to estimate uncertainties in the lepton momentum
scale and resolution, and correction factors and their associated
uncertainties are derived to match the simulated distributions to the observed
distributions~\cite{PERF-2013-05,PERF-2016-01,PERF-2014-05}.

\begin{figure}[htbp]
 	\centering
 	\includegraphics[width=0.45\textwidth]{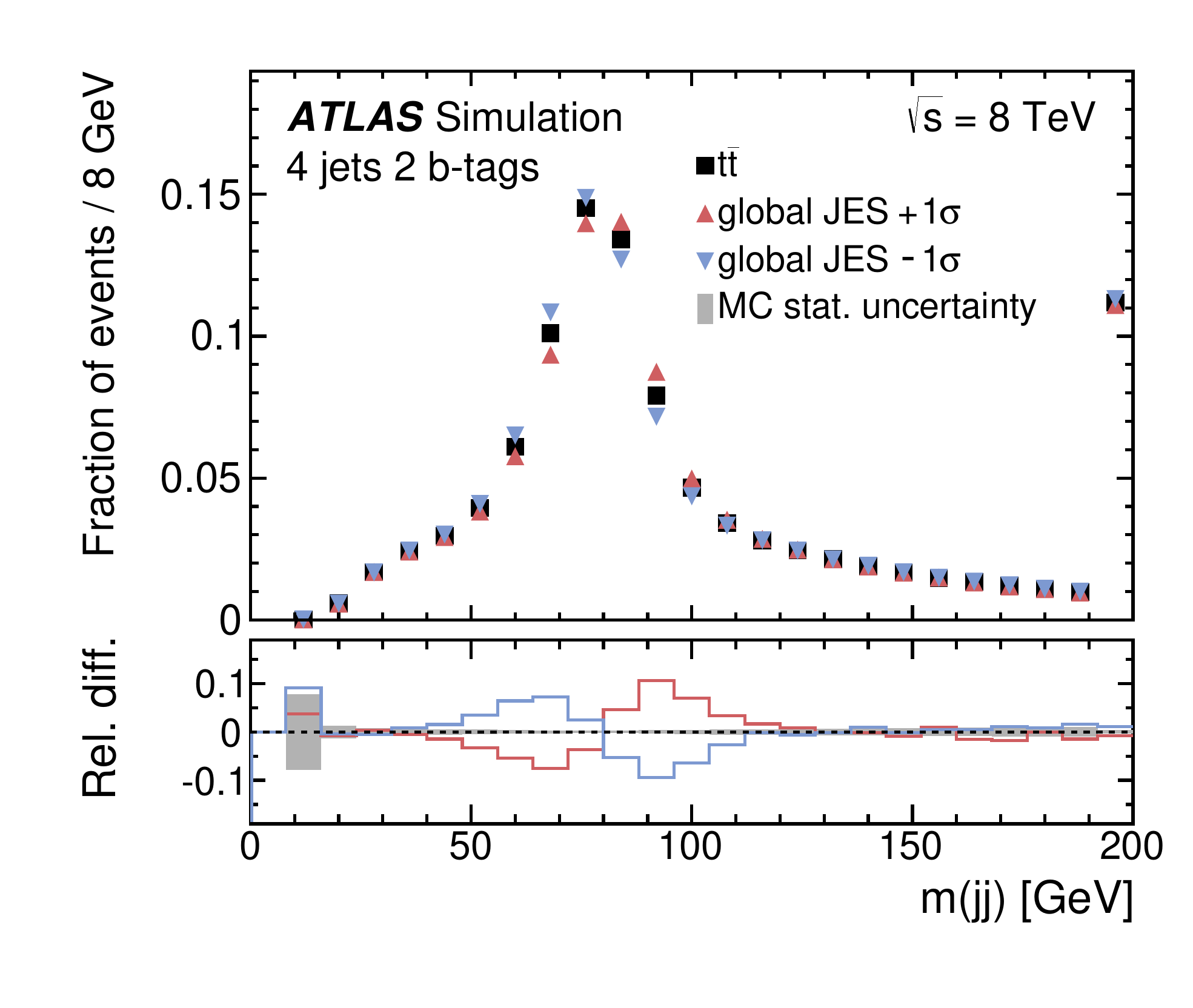}
 	\caption{ 
 	Probability density of the $m(jj)$ distribution from simulated \ttbar
 	events in \SRtwo for the nominal JES and the $\pm 1 \sigma$ variation. The
 	lower histogram shows the relative difference between the numbers of
 	\ttbar events for the $\pm 1 \sigma$ JES and the nominal JES in each bin with
 	respect to the nominal JES. The grey error band represents the statistical uncertainty of
 	the sample.
 	Events beyond the $x$-axis range are included in the last bin.}
 	\label{fig:shapesys_jes_mjj}
 \end{figure}

The JES uncertainties are derived using information from test-beam data,
collision data and simulation. The uncertainty is parameterised in terms of jet
\pt and $\eta$~\cite{PERF-2012-01}.
The JES uncertainty is broken down into various components originating from the
calibration method, the calorimeter response, the detector simulation, and the
set of parameters used in the MC event generator. Furthermore, contributions
from the modelling of pile-up effects, differences between jets induced by
$b$-quarks and those from gluons or light-quarks are included.
A large uncertainty in the JES originates from the a-priori unknown relative
fractions of quark-induced and gluon-induced jets in a generic sample, which is
normally assumed to be $(\num{50}\pm\num{50})\si{\percent}$. In this analysis,
the actual fraction of gluon-induced jets is estimated in simulated events, which leads to
a reduction in the uncertainty of these components by half. The fraction of
gluon-induced jets is obtained, considering all selected jets apart from
$b$-jets and it is between \SI{15}{\percent} to \SI{30}{\percent} depending on
the \pT and $\eta$ of the jet. The uncertainty in this fraction is estimated by
comparing different \ttbar samples, namely \POWHEGBOX + \PYTHIA, \POWHEGBOX +
\HERWIG, and \MCatNLO + \HERWIG as well as samples with varied scale settings in
the \POWHEGBOX + \PYTHIA set-up.
To estimate the systematic uncertainty of the JES, a parameterisation with
\num{25} uncorrelated components is used, as described in
Ref.~\cite{PERF-2012-01}.
For the purpose of the extraction of the \ttbar cross-section, a single
correction factor for the JES is included in the fit as a nuisance parameter
(see Sect.~\ref{sec:nn_stat}).
In this procedure, the dependence of the acceptance and the shape of the $m(jj)$
template distribution on the JES is parameterised using the global JES
uncertainty correction factor corresponding to the total JES uncertainty.
Figure~\ref{fig:shapesys_jes_mjj} shows the effect of a $\pm 1 \sigma$ change
in the global JES correction factor on the $m(jj)$ distribution.
When estimating the systematic uncertainty in the \ttbar cross-section due to
the JES in the statistical procedure, all \num{25} components are considered
and evaluated as described in Sect.~\ref{sec:nn_stat}.

Smaller uncertainties originate from modelling of the jet energy
resolution~\cite{ATLAS-CONF-2015-017,ATLAS-CONF-2015-057} and missing transverse
momentum~\cite{PERF-2014-04} to account for contributions from pile-up, soft
jets, and calorimeter cells not matched to any jets.
Uncertainties from the scale and 
resolution corrections for leptons and jets are propagated into the calculation 
of the missing transverse momentum as well.
The effect of uncertainties associated with the $\mathrm{JVF}$ is also considered for each jet.

Since the analysis makes use of $b$-tagging, the uncertainties in the
$b$-tagging efficiencies and the $c$-jet and light-jet mistag probabilities are
taken into account~\cite{ATLAS-CONF-2014-046,ATLAS-CONF-2014-004}. 
Correction factors applied to simulated events compensate for differences
between data and simulation in the tagging efficiency for $b$-jets, $c$-jets and
light-flavour jets. The correction for
$b$-jets
is derived from \ttbar
events in the dilepton channel and dijet events, and are found to be
consistent with unity with uncertainties at the level of a few percent over most
of the jet \pT range~\cite{ATLAS-CONF-2014-004}.
Similar to the JES, the uncertainty in the 
correction factor of the $b$-tagging efficiency is included as a nuisance
parameter in the fit for the extraction of the \ttbar cross-section. 
The parameterisation of the correction factor is obtained from the total
uncertainty in the $b$-tagging efficiency.

\subsection{Signal Monte Carlo modelling and parton distribution functions}
Systematic effects from MC modelling are estimated by comparing different
event generators and varying parameters for the event generation of the signal
process.

The uncertainty from renormalisation and factorisation scale variations, and
amount of additional radiation in the parton shower is estimated using the
\POWHEGBOX event generator interfaced to \PYTHIA by varying these scales and
using alternative sets of tuned parameters for the parton shower as described in
Sect.~\ref{sec:samples}.
Systematic effects due to the matching of the NLO matrix-element calculation and
the parton shower for \ttbar is estimated by comparing \MCatNLO with \POWHEGBOX,
both interfaced to the \HERWIG parton shower.
An uncertainty related to the modelling of parton-shower, hadronisation effects
and underlying-event, is estimated by comparing samples produced with \POWHEGBOX
+ \HERWIG  and \POWHEGBOX + \PYTHIA.
More details about these samples are given in Sect.~\ref{sec:samples}.

Systematic uncertainties related to the PDF set
are taken into account for the signal process.
The uncertainty is calculated following the PDF4LHC recommendation~\cite{Butterworth:2015oua}
using the \textsc{PDF4LHC15\_NLO} PDF set.
In addition, the acceptance difference between \textsc{PDF4LHC15\_NLO}
and \ct is considered, since the latter PDF set is not covered by the
uncertainty obtained with \textsc{PDF4LHC15\_NLO} and it is used in the simulation of \ttbar events.
This uncertainty is used in the final results, since it is larger than the
uncertainty obtained with \textsc{PDF4LHC15\_NLO}.

Finally, the statistical uncertainty of the MC samples as well as the \ZtoW data
sample is included.

\subsection{Background normalisation for non-fitted backgrounds}
Uncertainties in the normalisation of the non-fitted backgrounds, i.e.
single-top-quark, $VV$, and \Zjets events, are estimated using the uncertainties
in the theoretical cross-section predictions.
In the case of \Zjets, an uncertainty of \SI{24}{\percent} per additional jet is
added to the uncertainty of the inclusive cross-section in quadrature leading to
an total uncertainty of \SI{48}{\percent} for events with four jets. The uncertainty in
the multijet background is obtained in \SRone from the comparison between the fitting method and the \enquote{matrix} method as
detailed in Sect.~\ref{sec:nn_background}. For the other two regions, an uncertainty of
\SI{50}{\percent} is used.

\subsection{Background modelling}
Uncertainties in the shape of the \Wjets and multijet backgrounds are taken into
account for the discriminating observables used in the analysis. For the \Wjets
background, shape uncertainties are extracted from the differences between $Z$-boson and $W$-boson
production.
Although their production modes are very similar, differences exist in the
details of the production and decay. There are differences in heavy-flavour
production and in the helicity structures of the decay vertices.
Shape variations are built from a comparison of the NN discriminant and the
$m(jj)$ distribution between simulated \Wjets events, described in
Sect.~\ref{sec:samples}, and \ZtoW events derived from a simulated \Zjets
sample.
The uncertainty in the multijet background kinematics is estimated from the
differences between the predictions from the \enquote{jet-lepton} or \enquote{anti-muon} method 
and the \enquote{matrix} method in \SRoneNoSpace. 

\subsection{Luminosity}
The absolute luminosity scale is derived from beam-separation scans performed in November 2012. 
The uncertainty in the integrated luminosity is
\SI{1.9}{\percent}~\cite{DAPR-2013-01}.

\subsection{Beam energy}
The beam energy of the LHC was determined at \SI{4}{\TeV}
from the LHC magnetic model together with measurements
of the revolution frequency difference of proton and lead-ion beams,
with an uncertainty of \SI{0.1}{\percent}~\cite{PhysRevAccelBeams.20.081003}.
The impact of the uncertainty of the beam energy on the measured cross-section is negligible.

\section{Extraction of the \ttbar cross-section}
\label{sec:nn_stat}

The measured inclusive cross-section is given by
\begin{equation}
  \label{eq:inc_xs}
  \sigma_\mathrm{inc} = \frac{\hat{\nu}}{\epsilon \cdot
  \cal{L}_\mathrm{int}} = \frac{\hat{\beta} \cdot \nu}{\epsilon \cdot
  \cal{L}_\mathrm{int}} \quad \text{with} \quad \epsilon = \frac{N_\mathrm{sel}}{N_\mathrm{total}}\;,
\end{equation}
where $\hat{\nu}$ is the observed number of signal events.
The quantity $\epsilon$ is the total event-selection
efficiency, $N_\mathrm{total}$ is the number of events obtained from a
simulated signal sample before applying any requirement and $N_\mathrm{sel}$ is
the number of events obtained from the same simulated signal sample after applying
all selection requirements. In practice, $\hat{\nu}$ is given by $\hat{\beta}
\cdot \nu$, where $\hat{\beta}$ is an estimated scale factor obtained from a binned
maximum-likelihood fit and $\nu = \epsilon \cdot \sigma_{\mathrm{theo}} \cdot
\cal{L}_\mathrm{int}$ is the expected number of events for the signal process.
The reference cross-section $\sigma_{\mathrm{theo}}$ is defined by the
central value of the theoretical prediction given in Eq.~(\ref{eq:theo}). By
combining Eq.(\ref{eq:inc_xs}) with the expected number of events, one
obtains:
\begin{equation*}
  \sigma_\mathrm{inc} = \hat{\beta} \cdot \sigma_{\mathrm{theo}}.
\end{equation*}
The fiducial cross-section is given by
\begin{equation*}
  \sigma_\mathrm{fid} = A_\mathrm{fid}\cdot\sigma_\mathrm{inc} \quad
  \text{with} \quad  A_\mathrm{fid} = \frac{N_\mathrm{fid}}{N_\mathrm{total}}\;,
\end{equation*}
with $N_\mathrm{fid}$ being the number of events obtained from a simulated
signal sample after applying the particle-level selection. 
Here $A_\mathrm{fid}$ is defined for an inclusive \ttbar sample, including all
decay modes of the $W$ bosons. Using Eq.~(\ref{eq:inc_xs}), the fiducial
cross-section can be written as:
\begin{equation}
\label{eq:sigmafid_2}
  \sigma_\mathrm{fid} = \frac{\hat{\nu}}{\epsilon^{\prime} \cdot \cal{L}_\mathrm{int}}
  \quad \text{with} \quad \epsilon^{\prime} =
  \frac{N_\mathrm{sel}}{N_\mathrm{fid}} \;.
\end{equation}
From Eq.~(\ref{eq:sigmafid_2}) it is apparent that signal modelling
uncertainties that affect $N_\mathrm{sel}$ and $N_\mathrm{fid}$ in a
similar way give a reduced uncertainty in $\sigma_\mathrm{fid}$ compared to
that in $\sigma_\mathrm{inc}$.

The binned maximum-likelihood 
fit is performed simultaneously in the three signal regions defined in
Sect.~\ref{sec:sel}. 
For \SRone and \SRthr the distribution used in the fit is the NN
discriminant, while the invariant-mass distribution $m(jj)$ of the two untagged jets is used in \SRtwoNoSpace.
Electron- and muon-triggered events are combined in the templates used in this
fit.

Scale factors $\beta^{\ttbar}$ and $\beta^{W_j}$ for the signal and \Wjets
background, respectively, and two nuisance parameters $\delta_i$, namely the
$b$-tagging efficiency correction factor $\delta_{b-\mathrm{tag}}$ and the JES correction
factor $\delta_{\mathrm{JES}}$, are fitted in all three signal regions simultaneously.
The $\delta_i$ are defined such that \num{0} corresponds to the nominal value
and $\pm 1$ to a deviation of $\pm 1 \sigma$ of the corresponding systematic
uncertainty.

In order to account for differences in the flavour composition of the \Wjets
background, two uncorrelated scale factors are used: one in \SRone
($\beta^{W_1}$) and one in the two other signal regions ($\beta^{W_{2,3}}$).
The event yields of the
other backgrounds are not allowed to vary in the fit, but instead are
fixed to their predictions.
The likelihood function is given by the product of the Poisson
likelihoods in the individual bins $M$ of the histograms.
A Gaussian prior is incorporated into the likelihood function
to constrain $\delta_{b-\mathrm{tag}}$ within the associated
uncertainty:
\begin{eqnarray*}
L(\beta^{\ttbar}, \beta^{W_1},
\beta^{W_{2,3}} ,\delta_{b\text{-tag}},\delta_{\mathrm{JES}}) &=&
\prod_{k=1}^M
\frac{\mathrm{e}^{-\mu_k}\cdot\mu_k^{n_k}}{n_k!}
\;\cdot\; G(\delta_{b\text{-tag}}; 0, 1) 
\label{eq:Likelihood}
\end{eqnarray*}
with
\begin{equation*}
\begin{aligned}
\mu_k &=& \beta^s \cdot \nu_{s} \cdot \alpha^s_k + \sum_{j=1}^2
\beta^{W_j} \cdot \nu_{W_j} \cdot \alpha^{W_j}_k + \sum_{b=1}^4
\nu_{b} \cdot \alpha^{b}_k\ , \\
\beta^s &=& \beta^{\ttbar} \left\{1 + \sum_{i=1}^2 |\delta_i| \cdot\left(
H(\delta_i)\cdot \epsilon_{i+} + H(-\delta_i)\cdot \epsilon_{i-} \right)
\right\} \ ,  \\ 
\alpha^s_k &=& \alpha^{\ttbar}_k \sum_{i=1}^2 |\delta_i| \cdot\left\{
(\alpha_{ki}^+ - \alpha_{k})\cdot H(\delta_i) +
(\alpha_{ki}^- - \alpha_{k})\cdot H(-\delta_i)
\right\}.  \label{eq:shapesys}
\end{aligned}
\end{equation*}

The expected number of events $\mu_k$ in bin $k$ is the sum of the expected
number of events for the signal and the background processes. These are given by
the product of the predicted number of events $\nu_p$ of each process
and the fraction of events $\alpha^{p}_k$ in bin $k$ of the normalised
distribution.
Here $p$ denotes the signal $s$ and background processes $W_j$ and $b$, where
$b$ represents the background processes which are not varied in the fit.
The number of events observed in bin $k$ is denoted by $n_k$.
For the \ttbar signal, the scale factor $\beta^s$ contains 
the acceptance uncertainties for positive $\epsilon_{i+}$ and
negative $\epsilon_{i-}$ variations of the two profiled systematic
uncertainties, multiplied by their nuisance parameter $\delta_i$. 
The symbol $H$ denotes the Heaviside function.
The signal template shape for each profiled systematic variation
is calculated by interpolating in each bin $k$ between the standard
template $\alpha_{k}$ and the systematically altered histograms
$\alpha_{ki}^{\pm}$ using the nuisance parameter $\delta_i$ as a weight.
Linearity and closure tests are done to validate the statistical procedure.

The fit found the minimum of the negative log-likelihood 
function for the parameter values shown in Table~\ref{tab:betafit}.
The estimators for the nuisance parameters,  which parameterise their optimal
shift relative to the default value $0$ in terms of their uncertainty, are
found to be $\hat{\delta} $ = 0.62$\pm $ 0.09 for the $b$-tagging
efficiency correction factor and 0.68$\pm $0.07 for JES correction factor. 
This deviation of the $b$-tagging efficiency correction factor from the
nominal value of the simulated sample corresponds to a shift of the acceptance
in \SRone of \SI{1}{\percent} and \SI{2.6}{\percent} in \SRtwo and \SRthrNoSpace.
The deviation for the JES correction factor corresponds to a shift of the
acceptance of \SI{2.9}{\percent} in \SRoneNoSpace, of \SI{1.4}{\percent} in
\SRtwoNoSpace, and of \SI{4.4}{\percent} in \SRthrNoSpace. The deviation of
the JES correction factor also potentially accounts for differences in the
modelling of additional radiation between the different MC event generator
set-ups.
Finally, the fitted scale factor of the \Wjets process in \SRtwo and \SRthr
yields a value significantly higher than the one predicted by MC
simulation, consistent with previous measurements indicating an underestimate of
heavy-flavour production in the simulation~\cite{STDM-2012-11}.
 
\begin{table}
\sisetup{round-mode=figures, retain-explicit-plus}
\centering
\caption{\label{tab:betafit}Result of the maximum-likelihood fit to data. 
Estimators of the parameters of the likelihood function, the scale factor
$\hat{\beta}$ for the \ttbar and the two \Wjets channels 
and the derived contributions of the various processes to the three signal
regions are listed.
Only the statistical uncertainties obtained from the maximum-likelihood fit are
shown for \ttbar and \Wjets, while the normalisation uncertainties are
quoted for the other processes.} \begin{tabular}{lc
r@{$\,\pm\,$}S[table-format=4.0]
r@{$\,\pm\,$}S[table-format=4.0]
r@{$\,\pm\,$}S[table-format=4.0]
}
 \toprule
 Process          & $\hat{\beta}$  & \multicolumn{2}{c}{\SRone} & 
 \multicolumn{2}{c}{\SRtwo} & \multicolumn{2}{c}{\SRthr} \\
 \midrule
 \ttbar           &  $ 0.982 \pm 0.005 $  &  {\numRF{133388}{5}} & 630 & 
 {\numRF{64358}{4}} & 298 & {\numRF{62384}{4}} & 280  \\
 \midrule
 \Wjets 1 $b$-tag     & $ 1.08 \pm 0.02$   &  {\numRF{32154}{4}} & 477  &
 \multicolumn{2}{c}{--} & \multicolumn{2}{c}{--}  \\
 \Wjets $\ge$ 2 $b$-tags  & $ 1.41 \pm 0.08$   & \multicolumn{2}{c}{--} &
 {\numRF{3370}{3}} & 194 & {\numRF{2252}{3}} & 130  \\
 Single top       & -- &  {\numRF{11020}{4}} & 661 & {\numRF{3728}{3}} & 224
 & {\numRF{2593}{3}} & 156
 \\
 \Zjets           & -- &  {\numRF{3569}{2}}  & 1713  & {\numRF{406}{2}} &
 195  & 270 & 130  \\
 Diboson         	& -- &  {\numRF{1339}{2}} & 643  &  135 & 65  & 112 & 54 
 \\
 Multijet       	& -- &  {\numRF{10256}{3}} & 6872  & {\numRF{1943}{3}} &
 972  & {\numRF{1051}{3}} & 526 \\

 \midrule
 Total sum       & --   & {\numRF{191727}{4}}  & 7185 & {\numRF{73941}{3}}
 & 1078 & {\numRF{68662}{4}} & 645  \\
\hline
 Total observed  & --   & \multicolumn{1}{S[table-format=6.0,
 round-mode=off]@{\phantom{$\,\pm\,$}}}{192686}  & &
 \multicolumn{1}{S[table-format=5.0,
 round-mode=off]@{\phantom{$\,\pm\,$}}}{72978} & &
 \multicolumn{1}{S[table-format=5.0,
 round-mode=off]@{\phantom{$\,\pm\,$}}}{70120} \\
\bottomrule 
\end{tabular}
\end{table}

\begin{figure}[ht!]
\begin{center}
    \subfigure[]{
    \includegraphics[width=0.48\textwidth]{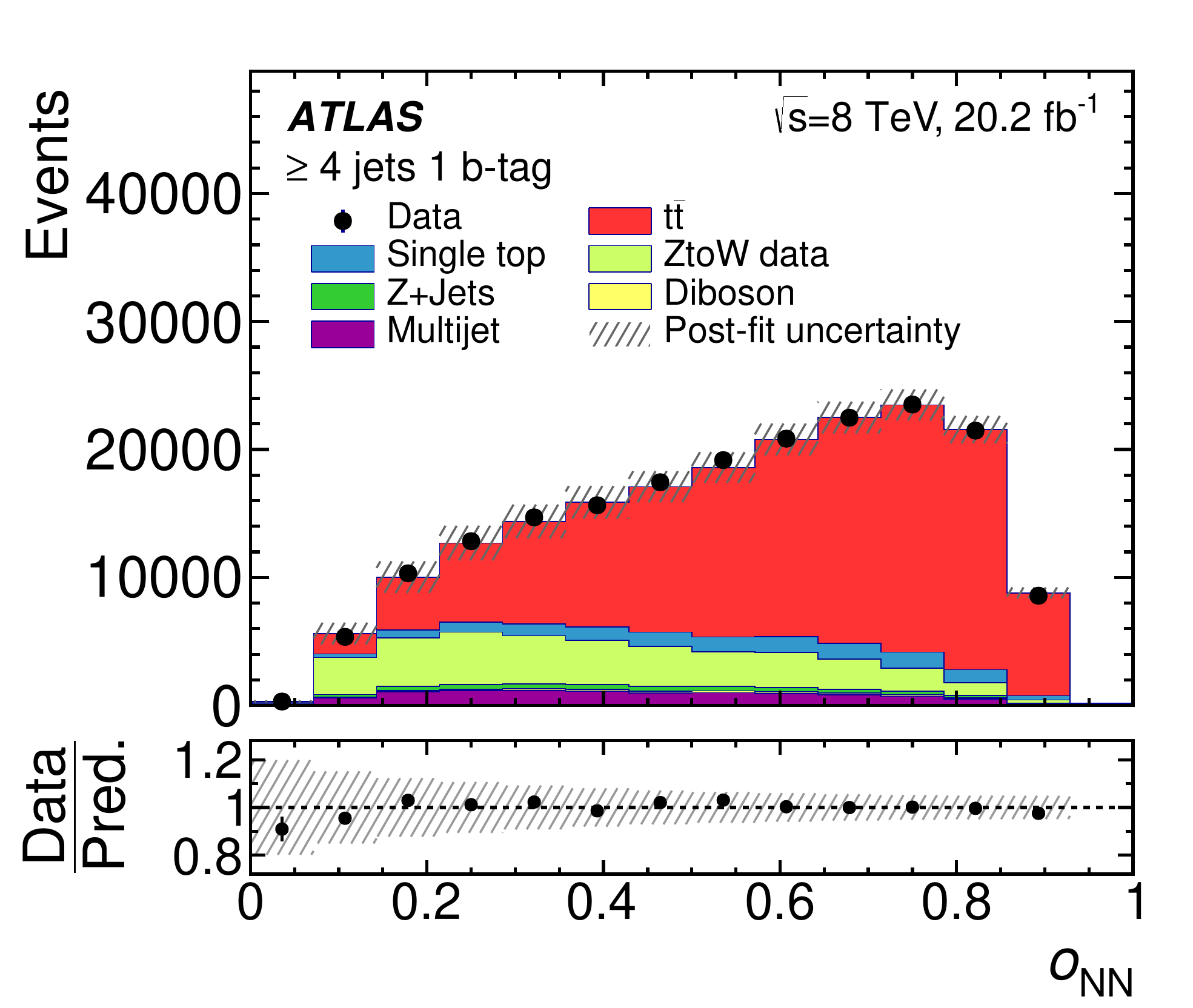}
    \label{subfig:postfit_nnnout_sr1}
    }
    \subfigure[]{
  \includegraphics[width=0.48\textwidth]{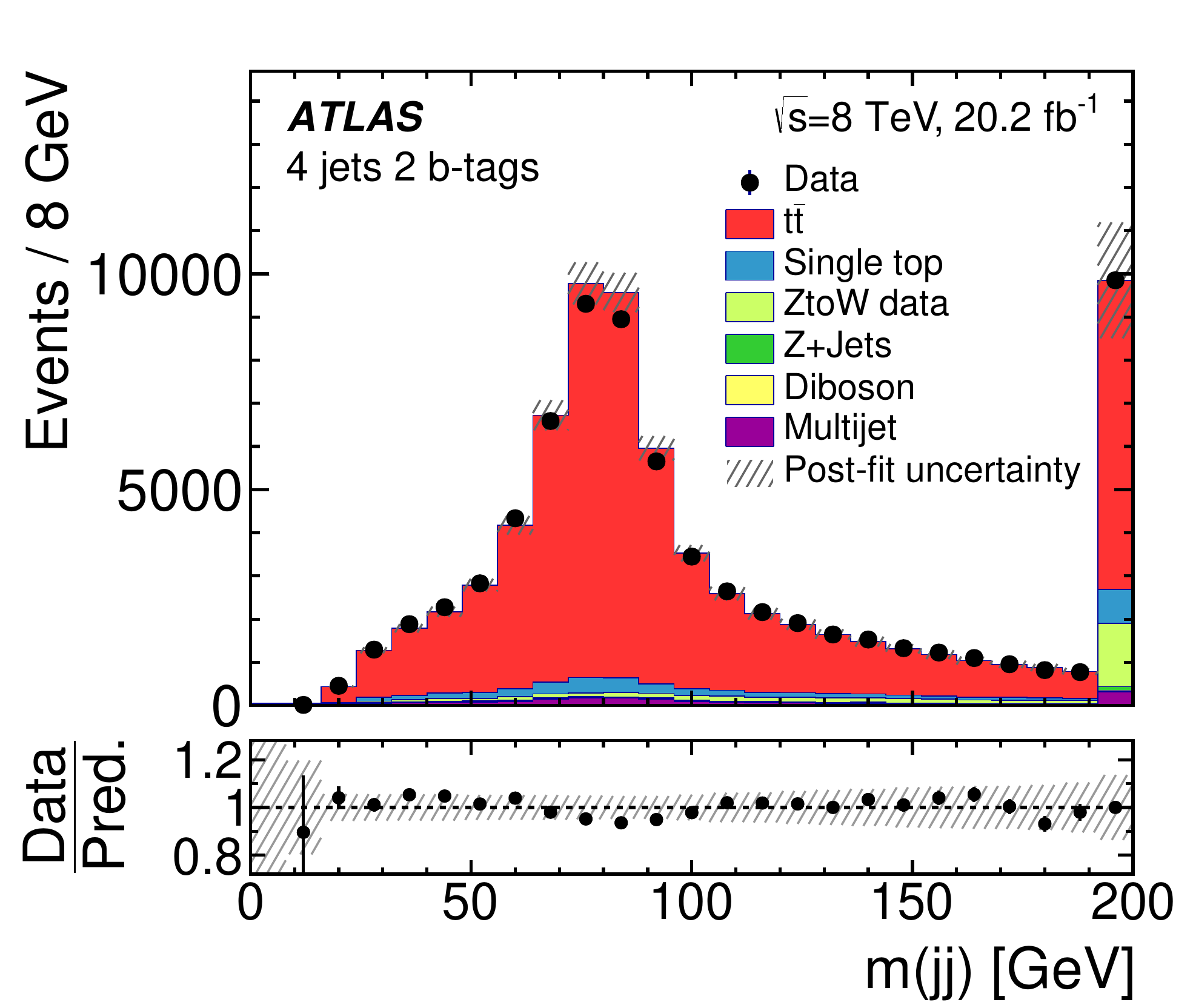}
  \label{subfig:postfit_mjj_sr2}
    }
    \subfigure[]{
  \includegraphics[width=0.48\textwidth]{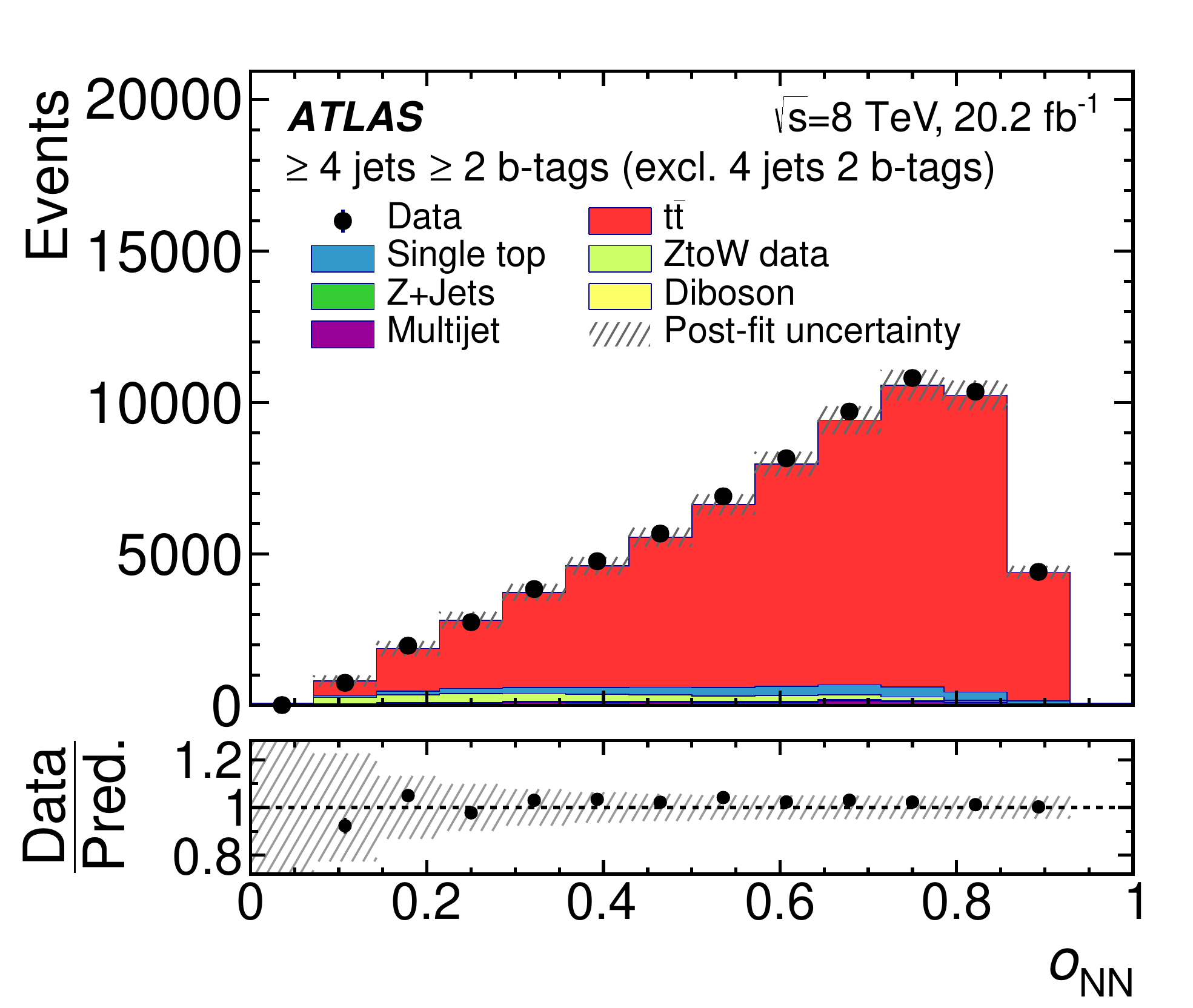}
  \label{subfig:postfit_nnnout_sr3}
    }
\caption{\label{fig:Fit}
Neural network discriminant $o_{\mathrm{NN}}$ or the $m(jj)$ distribution
normalised to the result of the maximum-likelihood fit for
\subref{subfig:postfit_nnnout_sr1} \SRoneNoSpace,
\subref{subfig:postfit_mjj_sr2} \SRtwoNoSpace, and
\subref{subfig:postfit_nnnout_sr3} \SRthrNoSpace.
     The hatched error bands represent the post-fit uncertainty.
     The ratio of observed to predicted (Pred.) number of events in each bin is
     shown in the lower histogram. Events beyond the $x$-axis range are included
     in the last bin.}
\end{center}
\end{figure}

\begin{figure}[htbp]
    \centering
    \subfigure[]{
	\includegraphics[width=0.42\textwidth]{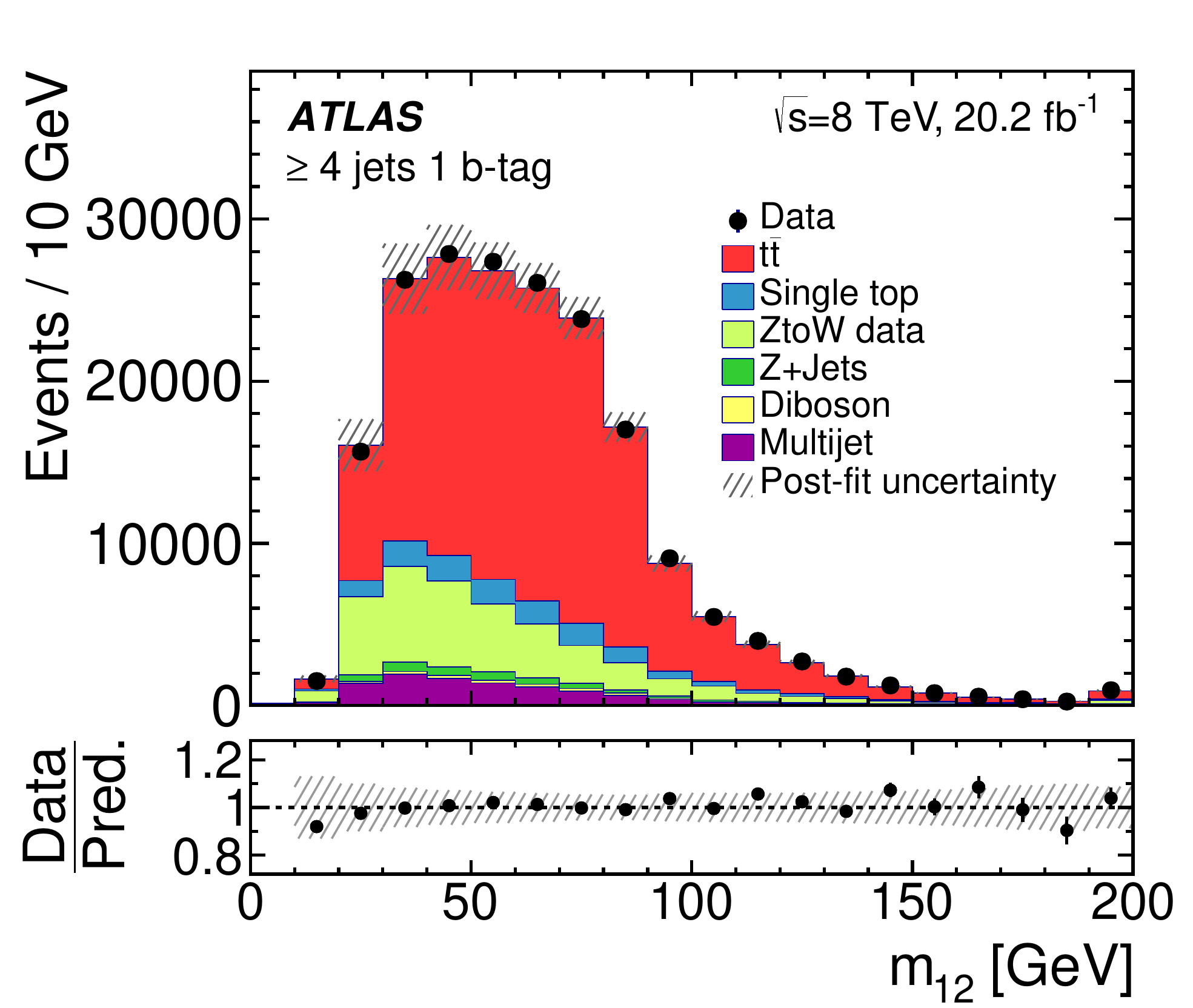} 
	\label{subfig:postfit_m12_ch1}
	}
	\subfigure[]{
	\includegraphics[width=0.42\textwidth]{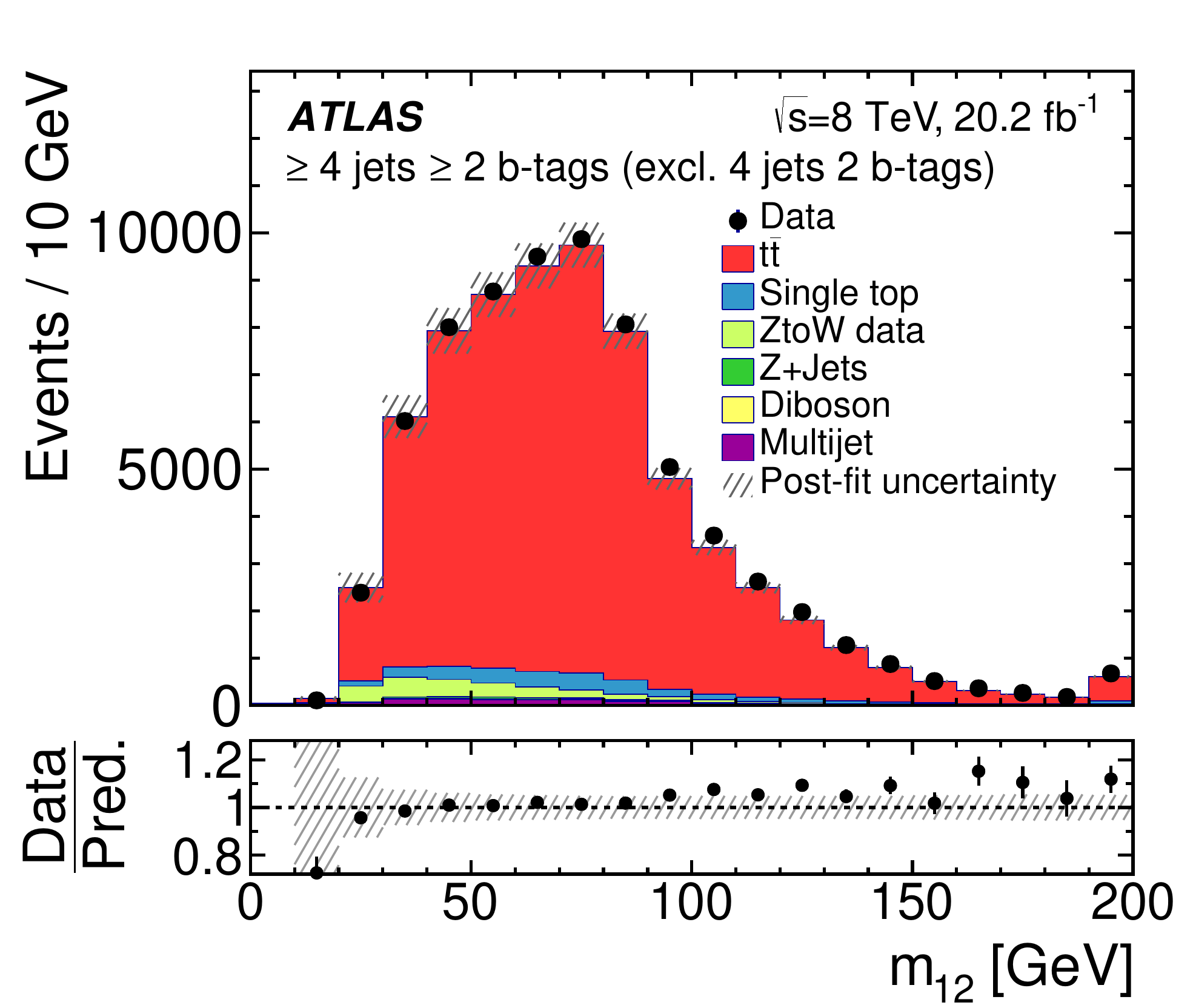} 
	\label{subfig:postfit_m12_ch3}
	}
	\subfigure[]{
	\includegraphics[width=0.42\textwidth]{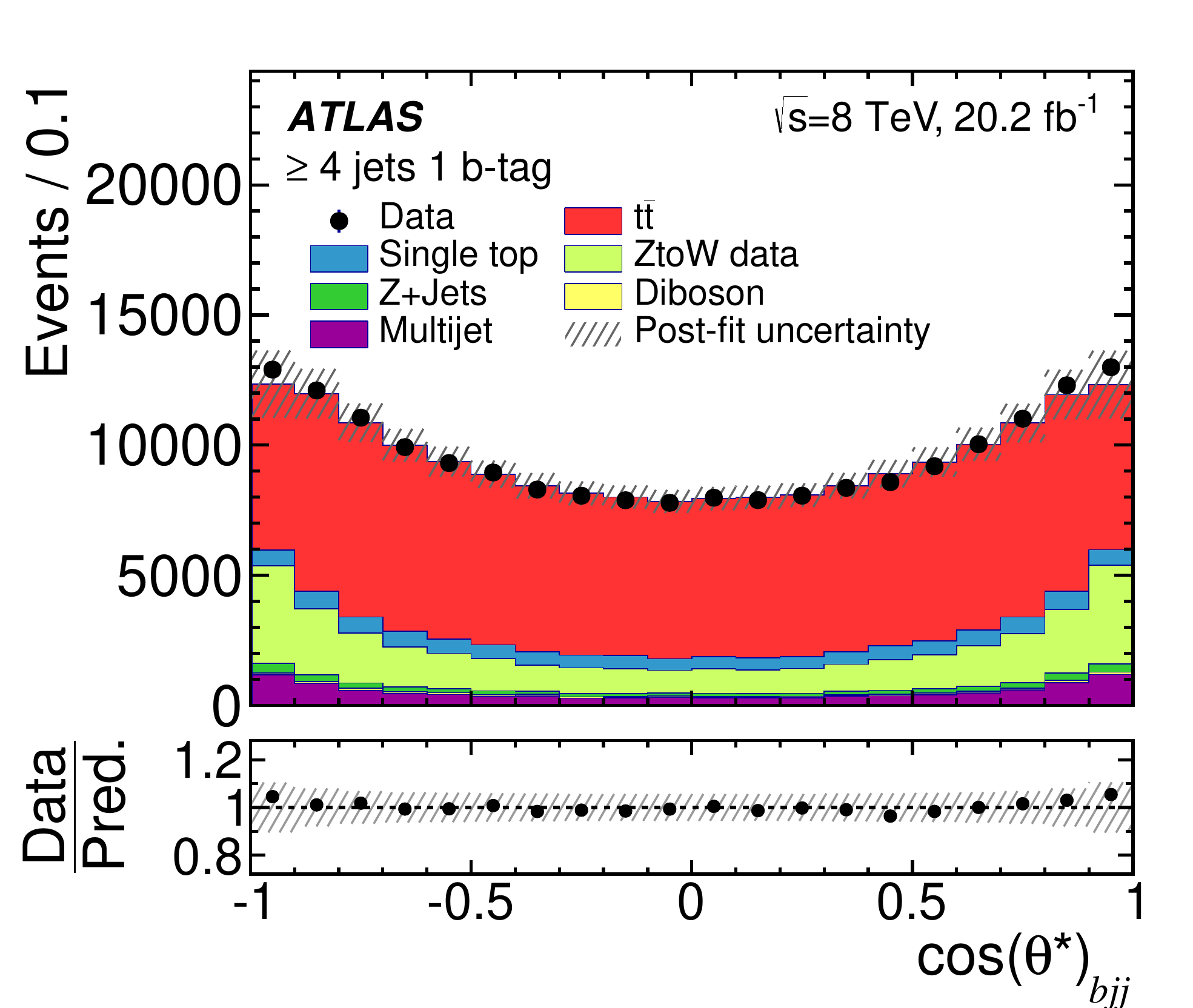}
	\label{subfig:postfit_costheta_ch1}
	}	
	\subfigure[]{
	\includegraphics[width=0.42\textwidth]{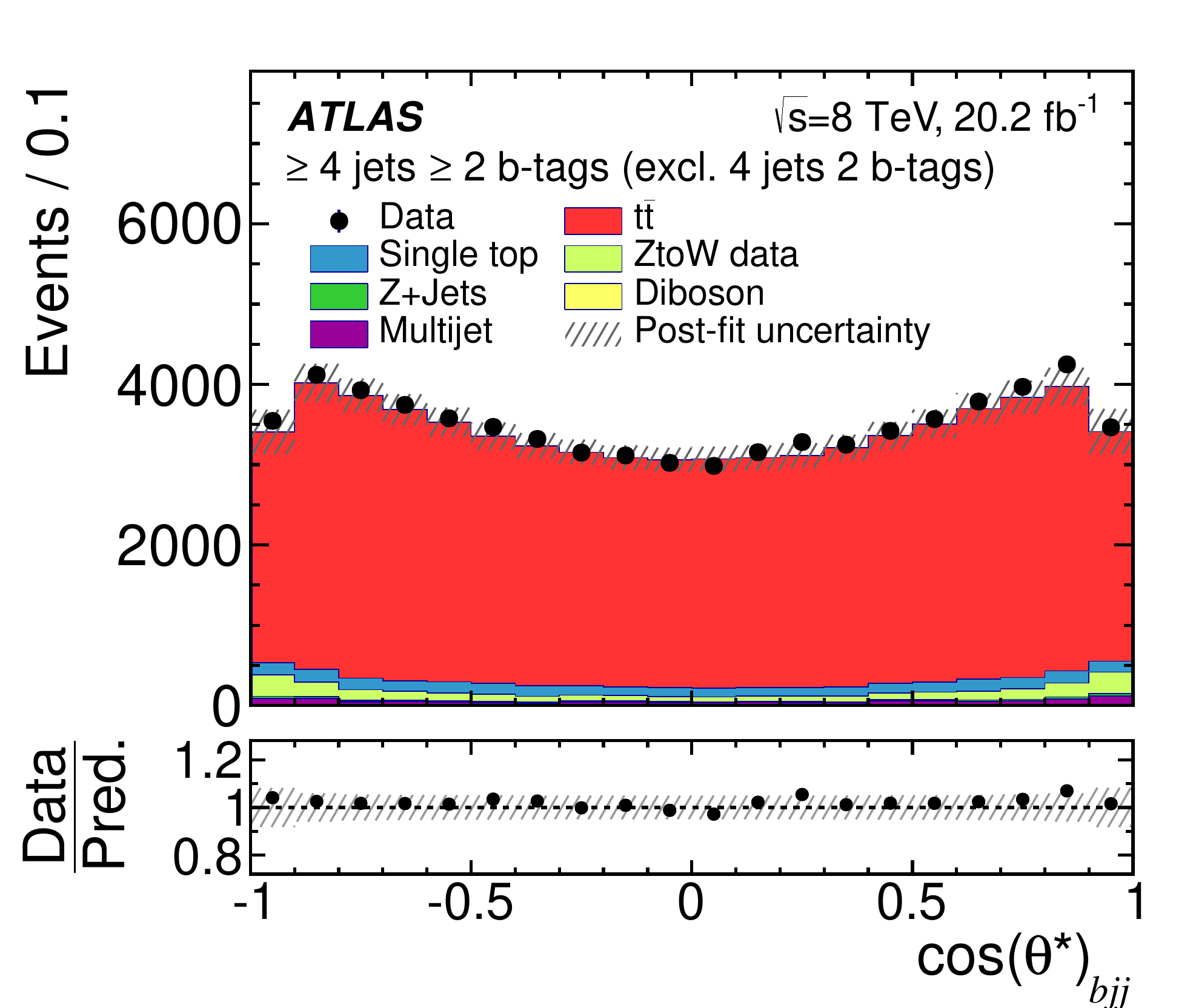}
	\label{subfig:postfit_costheta_ch3}
	}
	\subfigure[]{
	\includegraphics[width=0.42\textwidth]{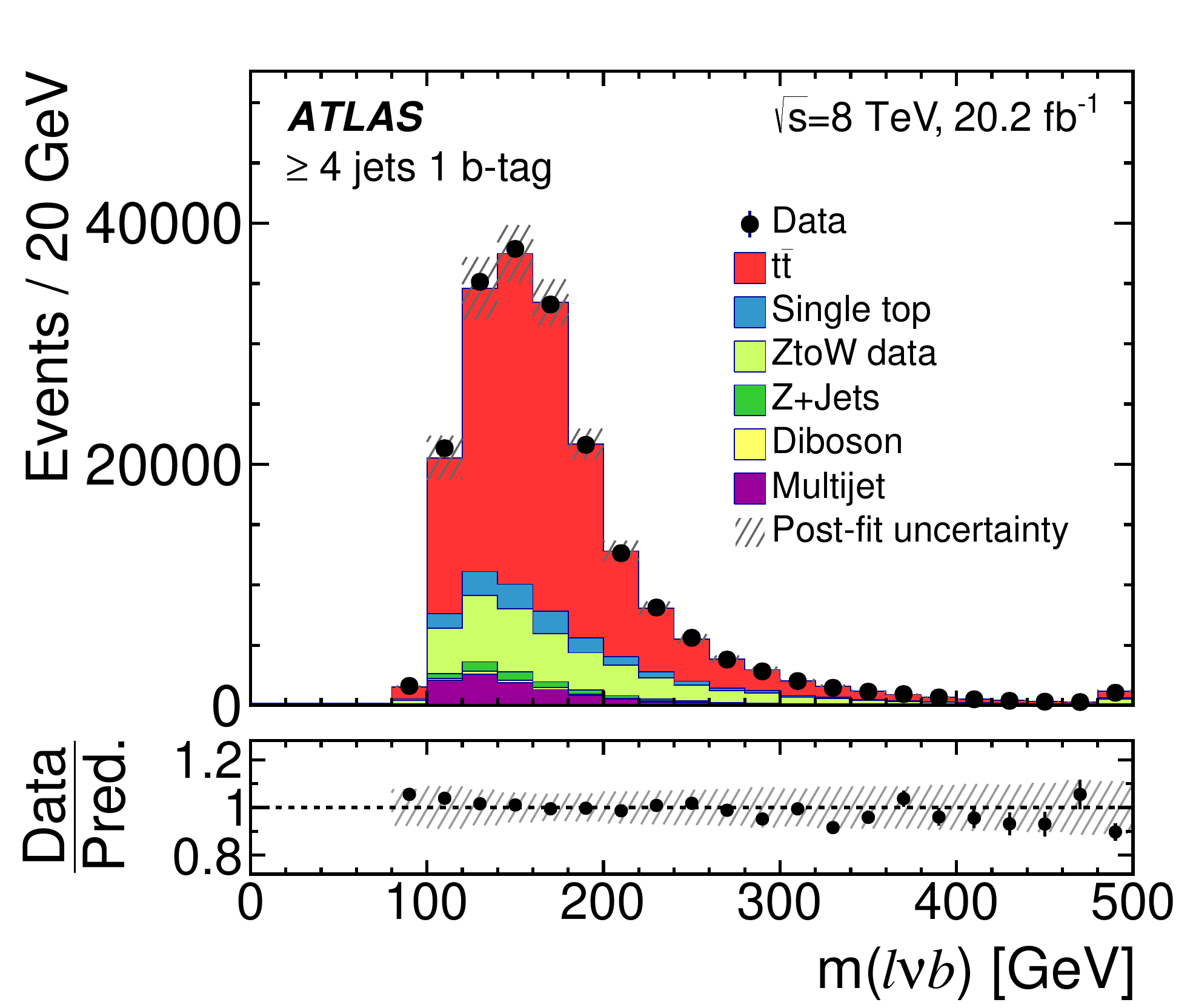} 
	\label{subfig:postfit_mlnub_ch1}
	}	
	\subfigure[]{
	\includegraphics[width=0.42\textwidth]{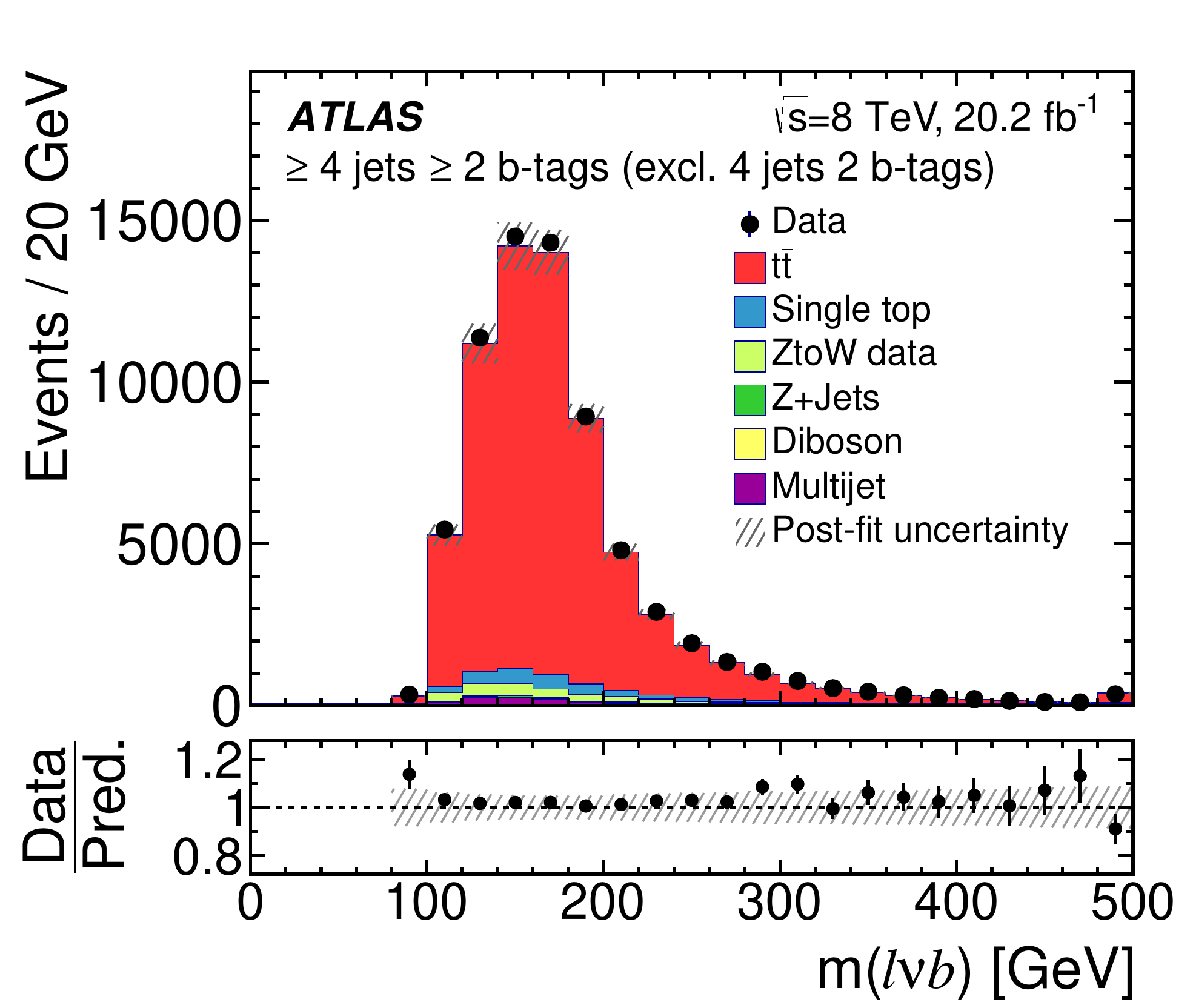} 
	\label{subfig:postfit_mlnub_ch3}
	}
    \caption{Distributions of the three most discriminating NN input variables
    for (left) \SRone and (right) \SRthrNoSpace. The signal and backgrounds are
    normalised to the result of the maximum-likelihood
    fit: \subref{subfig:postfit_m12_ch1}-\subref{subfig:postfit_m12_ch3}
    smallest invariant mass between jet pairs,
    \subref{subfig:postfit_costheta_ch1}-\subref{subfig:postfit_costheta_ch3}
    cosine of the angle between the hadronic-top-quark momentum and the beam
    direction in the \ttbar rest frame, and \subref{subfig:postfit_mlnub_ch1}-\subref{subfig:postfit_mlnub_ch3}
    mass of the reconstructed semileptonically decaying top quark.
    The hatched error bands represent the
    post-fit uncertainty. The ratio of observed to predicted (Pred.) number of
    events in each bin is shown in the lower histogram. Events beyond the
    $x$-axis range are included in the last bin.}
    \label{fig:postfit-vars}
\end{figure}

The signal and background templates scaled and morphed to the fitted
values of the fit parameters are compared to the observed distributions of the
NN discriminant in \SRone and \SRthr and the $m(jj)$ distribution in \SRtwoNoSpace, shown in
Fig.~\ref{fig:Fit}. Comparisons of the data and the fit results are shown for
the three most discriminating input variables of the NN in
Fig.~\ref{fig:postfit-vars} for \SRone and for \SRthrNoSpace.

The systematic uncertainties in the cross-section measurements are estimated
using pseudo-experiments.
In each of these experiments, the detector effects, background contributions
and model uncertainties are varied within their systematic uncertainties. They impact the yields 
of the processes and shapes of the template distributions used to create the
pseudo-datasets in the three signal regions.
Correlations between rate and shape uncertainties for a given component are
taken into account.
The entire set of pseudo-experiments can thereby be interpreted as a replication
of the sample space of all systematic variations. By measuring the \ttbar
cross-section, an estimator of the probability density of all possible outcomes
of the measurement is obtained. The RMS of this estimator distribution is itself
an estimator of the observed uncertainties.
Using the measured \ttbar cross-section and the estimated nuisance parameters,
the uncertainty of the actual measurement is estimated.

The total uncertainty in both the inclusive and the fiducial \ttbar
cross-section is presented in Table~\ref{tab:observedSys} and is estimated to be
$\SI{5.7}{\percent}$ for the inclusive measurement and $\SI{4.5}{\percent}$ for
the fiducial measurement.
The breakdown of the contributions from individual, or categories of, systematic
uncertainties are also listed. In this case, only the considered source or group
of sources is varied in the generation of the pseudo-experiments. The largest
uncertainty in the inclusive measurement is due to the uncertainty in the PDF
sets and the MC modelling of the signal process. The effects of uncertainties
in the JES and the $b$-tagging efficiency have been significantly reduced by
including them as nuisance parameters together with the chosen signal regions
and discriminant distributions.
Furthermore, the uncertainty due to additional radiation is reduced by a factor
of three thanks to the inclusion of the $m(jj)$ distribution in the analysis.
For the fiducial cross-section measurement,
the uncertainties in the MC modelling and PDF sets are reduced. The uncertainty
in the \ttbar cross-section due to the PDF sets is largest for events which are
produced in the forward direction, i.e. one initial gluon has a high Bjorken-$x$ value.
The PDFs for high-$x$ gluons have large uncertainties in all current PDF sets.
Selecting events in the fiducial volume reduces the fraction of such events
significantly and therefore the uncertainty is reduced significantly as well.
In the case of the MC modelling, the uncertainty due to additional radiation is
reduced more than the parton-shower and NLO-matching uncertainties, since
varying the amount of radiation leads to similar changes in the selection
efficiencies of the fiducial and reconstructed volumes and therefore to smaller
uncertainties in the \ttbar cross-section.

\begin{table}[ht!]
\caption{\label{tab:observedSys} Breakdown of relative uncertainties in
the measured inclusive and fiducial \ttbar cross-sections. The total
uncertainties contain all considered uncertainties.}
\begin{center}
\begin{tabular}{lrr}
\toprule
Source & \multicolumn{1}{c}{$\frac{\Delta
\sigma_{\text{inc}}}{\sigma_{\text{inc}}}$ [\%]} &
\multicolumn{1}{c}{$\frac{\Delta \sigma_{\text{fid}}}{\sigma_{\text{fid}}}$ [\%]}
\\
\midrule
Statistical uncertainty & \num{0.3} & \num{0.3}\\
\midrule
\textbf{Physics object modelling}  &   \\
Jet energy scale & \num{1.1} & \num{1.1} \\
Jet energy resolution & \num{0.1} & \num{0.1}  \\
Jet reconstruction efficiency & \num{< 0.1} & \num{< 0.1}  \\
\MET scale & \num{0.1} & \num{0.1} \\
\MET resolution & \num{< 0.1} & \num{< 0.1} \\
Muon momentum scale & \num{< 0.1} & \num{< 0.1} \\
Muon momentum resolution & \num{< 0.1} & \num{< 0.1} \\
Electron energy scale & \num{0.1} & \num{0.1} \\
Electron energy resolution & \num{< 0.1} & \num{< 0.1}  \\
Lepton identification & \num{1.4} & \num{1.4} \\
Lepton reconstruction & \num{0.3}& \num{0.3}  \\
Lepton trigger & \num{1.3}  & \num{1.3} \\
$b$-tagging efficiency & \num{0.3} & \num{0.3} \\
$c$-tagging efficiency & \num{0.5} & \num{0.5}  \\
Mistag rate & \num{0.3} & \num{0.3} \\
\midrule
\textbf{Signal Monte Carlo modelling and parton distribution functions}  &    \\
NLO matching & \num{1.1} & \num{0.9} \\
Scale variations & \num{2.2} & \num{1.0}  \\
Parton shower & \num{1.3} & \num{0.9}  \\
PDF & \num{3.0} & \num{0.1} \\
\midrule
\textbf{Background normalisation for non-fitted backgrounds}  &   \\
Single top & \num{0.3} & \num{0.3} \\
\Zjets & \num{0.2} & \num{0.2} \\
Diboson & \num{0.1} & \num{0.1} \\
\midrule
\textbf{Background modelling} & \\
\ZtoW modelling & \num{1.1} & \num{1.1}  \\
Multijet & \num{0.6} & \num{0.6} \\
\midrule
\textbf{Luminosity} & \num{1.9} & \num{1.9} \\ 
\midrule
\textbf{Total~(syst.)} & \num{5.7} & \num{4.5}\\
\midrule
\textbf{Total~(syst.+stat.)} & \num{5.7} & \num{4.5} \\
\bottomrule
\end{tabular}
\end{center}
\end{table}

\clearpage
\section{Results}
\label{sec:nn_result}
After performing a binned maximum-likelihood fit to the NN discriminant
distributions and the $m(jj)$ distribution, and estimating the total uncertainty, the inclusive
\ttbar cross-section is measured to be:
\begin{eqnarray*}
\sigma_{\text{inc}}(\ttbar) = 248.3 \pm 0.7 \, (\mathrm{stat.})  \pm 13.4 \,
(\mathrm{syst.}) \pm 4.7 \, (\mathrm{lumi.}) \ \mathrm{pb} 
\end{eqnarray*}
assuming a top-quark mass of $\mtop = \SI{172.5}{\gev}$.

The fiducial cross-section measured in the fiducial volume defined in
Sect.~\ref{sec:fid} with acceptance $A_{\text{fid}}=\SI{19.6}{\percent}$
is:
\begin{eqnarray*}
\sigma_{\text{fid}}(\ttbar) = 48.8 \pm 0.1 \, (\mathrm{stat.})  \pm 2.0 \,
(\mathrm{syst.}) \pm 0.9 \, (\mathrm{lumi.})~\text{pb}.
\end{eqnarray*}
The dependence of the inclusive \ttbar cross-section measurement on the assumed
value of $\mtop$ is mainly due to acceptance effects and can be expressed by the function:
\begin{equation*}
\sigma_{\ttbar}(\mtop) = \sigma_{\ttbar}(\SI{172.5}{\gev}) + p_1 \cdot \Delta
\mtop  + p_2 \cdot \Delta \mtop^{2}\;,
\end{equation*}
with $\Delta \mtop = \mtop - \SI{172.5}{\gev}$. 
The parameters $p_1=-2.07 \pm \SI{0.07}{\pb / \gev}$ and $p_2=0.07 \pm
\SI{0.02}{\pb / \gev^2}$ are determined using dedicated signal samples with
different $\mtop$ values, where signal template distributions are
obtained from the alternative samples and the fit to data is repeated.

A combination of the cross-section in this channel with the more
precise result in the dilepton channel~\cite{TOPQ-2013-04} was tested.
The central values of the two results are consistent within \SI{0.2}{\percent},
but due to the higher precision of the dilepton result, the combination yielded
only a marginal improvement.

\FloatBarrier


\section{Conclusions}
\label{sec:conclusion}

A measurement of both the inclusive and fiducial \ttbar cross-sections in $pp$
collisions at $\sqrt{s} = \SI{8}{\TeV}$ in the lepton+jets channel is presented
using data collected in 2012 with the ATLAS detector at the LHC, corresponding
to an integrated luminosity of \SI{20.2}{\per\fb}.

In order to reduce major uncertainties coming from the jet energy scale and the
$b$-tagging efficiency, the analysis splits the selected data sample into three
disjoint signal regions with different numbers of $b$-tagged jets and different
jet multiplicities.
Using an artificial neural network, the separation between the signal and
background processes is improved compared to using single observables.
Additionally, the analysis makes use of a data-driven approach to model the
dominant \Wjets background. It is modelled from collision data by
converting \Zjets candidate events into a \Wjets sample.

The \ttbar cross-section is determined using a binned maximum-likelihood fit to
the three signal regions, constraining correction factors for the jet
energy scale and the $b$-tagging efficiency.
The inclusive \ttbar cross-section is measured with a precision of
\SI{5.7}{\percent} to be:
\begin{eqnarray*}
\sigma_{\text{inc}}(\ttbar) = 248.3 \pm 0.7 \, (\mathrm{stat.})  \pm 13.4 \,
(\mathrm{syst.}) \pm 4.7 \, (\mathrm{lumi.}) \ \mathrm{pb} 
\end{eqnarray*}
assuming a top-quark mass of $\mtop = \SI{172.5}{\gev}$.

The fiducial cross-section is measured with a precision of \SI{4.5}{\percent} to
be:
\begin{eqnarray*}
\sigma_{\text{fid}}(\ttbar) = 48.8 \pm 0.1 \, (\mathrm{stat.})  \pm 2.0 \,
(\mathrm{syst.}) \pm 0.9 \, (\mathrm{lumi.})~\text{pb}.
\end{eqnarray*}
This result is a significant improvement on the previous ATLAS measurement at
$\sqrt{s} = \SI{8}{\TeV}$ in the lepton+jets channel and is in agreement with
measurements of the inclusive \ttbar cross-section in other decay modes and with
the theoretical prediction.


\section*{Acknowledgements}


We thank CERN for the very successful operation of the LHC, as well as the
support staff from our institutions without whom ATLAS could not be
operated efficiently.

We acknowledge the support of ANPCyT, Argentina; YerPhI, Armenia; ARC, Australia; BMWFW and FWF, Austria; ANAS, Azerbaijan; SSTC, Belarus; CNPq and FAPESP, Brazil; NSERC, NRC and CFI, Canada; CERN; CONICYT, Chile; CAS, MOST and NSFC, China; COLCIENCIAS, Colombia; MSMT CR, MPO CR and VSC CR, Czech Republic; DNRF and DNSRC, Denmark; IN2P3-CNRS, CEA-DRF/IRFU, France; SRNSFG, Georgia; BMBF, HGF, and MPG, Germany; GSRT, Greece; RGC, Hong Kong SAR, China; ISF, I-CORE and Benoziyo Center, Israel; INFN, Italy; MEXT and JSPS, Japan; CNRST, Morocco; NWO, Netherlands; RCN, Norway; MNiSW and NCN, Poland; FCT, Portugal; MNE/IFA, Romania; MES of Russia and NRC KI, Russian Federation; JINR; MESTD, Serbia; MSSR, Slovakia; ARRS and MIZ\v{S}, Slovenia; DST/NRF, South Africa; MINECO, Spain; SRC and Wallenberg Foundation, Sweden; SERI, SNSF and Cantons of Bern and Geneva, Switzerland; MOST, Taiwan; TAEK, Turkey; STFC, United Kingdom; DOE and NSF, United States of America. In addition, individual groups and members have received support from BCKDF, the Canada Council, CANARIE, CRC, Compute Canada, FQRNT, and the Ontario Innovation Trust, Canada; EPLANET, ERC, ERDF, FP7, Horizon 2020 and Marie Sk{\l}odowska-Curie Actions, European Union; Investissements d'Avenir Labex and Idex, ANR, R{\'e}gion Auvergne and Fondation Partager le Savoir, France; DFG and AvH Foundation, Germany; Herakleitos, Thales and Aristeia programmes co-financed by EU-ESF and the Greek NSRF; BSF, GIF and Minerva, Israel; BRF, Norway; CERCA Programme Generalitat de Catalunya, Generalitat Valenciana, Spain; the Royal Society and Leverhulme Trust, United Kingdom.

The crucial computing support from all WLCG partners is acknowledged gratefully, in particular from CERN, the ATLAS Tier-1 facilities at TRIUMF (Canada), NDGF (Denmark, Norway, Sweden), CC-IN2P3 (France), KIT/GridKA (Germany), INFN-CNAF (Italy), NL-T1 (Netherlands), PIC (Spain), ASGC (Taiwan), RAL (UK) and BNL (USA), the Tier-2 facilities worldwide and large non-WLCG resource providers. Major contributors of computing resources are listed in Ref.~\cite{ATL-GEN-PUB-2016-002}.

\printbibliography


\clearpage 
\begin{flushleft}
{\Large The ATLAS Collaboration}

\bigskip

M.~Aaboud$^\textrm{\scriptsize 137d}$,
G.~Aad$^\textrm{\scriptsize 88}$,
B.~Abbott$^\textrm{\scriptsize 115}$,
O.~Abdinov$^\textrm{\scriptsize 12}$$^{,*}$,
B.~Abeloos$^\textrm{\scriptsize 119}$,
S.H.~Abidi$^\textrm{\scriptsize 161}$,
O.S.~AbouZeid$^\textrm{\scriptsize 139}$,
N.L.~Abraham$^\textrm{\scriptsize 151}$,
H.~Abramowicz$^\textrm{\scriptsize 155}$,
H.~Abreu$^\textrm{\scriptsize 154}$
% ...
\bigskip
\\
% ...
$^{12}$ Institute of Physics, Azerbaijan Academy of Sciences, Baku, Azerbaijan\\
% ...
$^{88}$ CPPM, Aix-Marseille Universit{\'e} and CNRS/IN2P3, Marseille, France\\
% ...
$^{115}$ Homer L. Dodge Department of Physics and Astronomy, University of Oklahoma, Norman OK, United States of America\\
% ...
$^{119}$ LAL, Univ. Paris-Sud, CNRS/IN2P3, Universit{\'e} Paris-Saclay, Orsay, France\\
% ...
$^{137}$ $^{(a)}$ Facult{\'e} des Sciences Ain Chock, R{\'e}seau Universitaire de Physique des Hautes Energies - Universit{\'e} Hassan II, Casablanca; $^{(b)}$ Centre National de l'Energie des Sciences Techniques Nucleaires, Rabat; $^{(c)}$ Facult{\'e} des Sciences Semlalia, Universit{\'e} Cadi Ayyad, LPHEA-Marrakech; $^{(d)}$ Facult{\'e} des Sciences, Universit{\'e} Mohamed Premier and LPTPM, Oujda; $^{(e)}$ Facult{\'e} des sciences, Universit{\'e} Mohammed V, Rabat, Morocco\\
% ...
$^{139}$ Santa Cruz Institute for Particle Physics, University of California Santa Cruz, Santa Cruz CA, United States of America\\
% ...
$^{151}$ Department of Physics and Astronomy, University of Sussex, Brighton, United Kingdom\\
% ...
$^{154}$ Department of Physics, Technion: Israel Institute of Technology, Haifa, Israel\\
$^{155}$ Raymond and Beverly Sackler School of Physics and Astronomy, Tel Aviv University, Tel Aviv, Israel\\
% ...
$^{161}$ Department of Physics, University of Toronto, Toronto ON, Canada\\
% ...
$^{a}$ Also at Department of Physics, King's College London, London, United Kingdom\\
$^{b}$ Also at Institute of Physics, Azerbaijan Academy of Sciences, Baku, Azerbaijan\\
$^{c}$ Also at Novosibirsk State University, Novosibirsk, Russia\\
$^{d}$ Also at TRIUMF, Vancouver BC, Canada\\
$^{e}$ Also at Department of Physics {\&} Astronomy, University of Louisville, Louisville, KY, United States of America\\
% ...
$^{*}$ Deceased
\end{flushleft}


%
\FloatBarrier

\end{document}